\documentclass[12pt,final]{elsart}

  \newcommand{\preprint}{SACLAY--T07/013\\BNL--NT--07/10}

  \usepackage{latexsym,bm,amsmath,amssymb,amsfonts}
  \usepackage{epsfig,graphics,graphicx}
  \usepackage{cite}
  \usepackage{slashed}

  \long\def\comment#1{ }

  \newcommand{\dif}{{\rm d}}

  \newcommand{\del}{\partial}
  \newcommand{\lan}{\left\langle}
  \newcommand{\ran}{\right\rangle}
  \newcommand{\mcal}{\mathcal}
  \newcommand{\rme}{{\rm e}}
  \newcommand{\rmi}{{\rm i}}

  \newcommand{\grad}{\nabla}

  \newcommand{\nn}{\nonumber\\}

  \newcommand{\beq}{\begin{eqnarray}}
  \newcommand{\eeq}{\end{eqnarray}}
  
 \def\simge{\mathrel{%
   \rlap{\raise 0.511ex \hbox{$>$}}{\lower 0.511ex \hbox{$\sim$}}}}
\def\simle{\mathrel{
   \rlap{\raise 0.511ex \hbox{$<$}}{\lower 0.511ex \hbox{$\sim$}}}}

\begin{document}

\begin{flushright}
{\small \preprint}
\end{flushright}
\vspace{1.5cm}

\begin{frontmatter}\parbox[]{16.0cm}{\begin{center}
\title{\rm \LARGE Liouville field
theory for gluon saturation\\ in QCD at high energy}

\author{E.~Iancu$^{\rm a,1,2}$}%\thanksref{th2}},
\author{ and L.~McLerran$^{\rm b,c,1}$}

\address{$^{\rm a}$ Service de Physique Theorique, CEA Saclay,
CEA/DSM/SPhT, F-91191 Gif-sur-Yvette, France}
\address{$^{\rm b}$ RIKEN BNL Research Center, Brookhaven National Laboratory,
Upton, NY 11973, USA}
\address{$^{\rm c}$ Physics Department, Brookhaven
National Laboratory, Upton, NY 11973, USA}

\thanks{{\it E-mail addresses:}
iancu@dsm-mail.cea.fr (E.~Iancu), mclerran@quark.phy.bnl.gov
(L.~McLerran).}
\thanks{Membre du Centre National de la Recherche Scientifique
(CNRS), France.}

{\small \today}
%\vspace{0.8cm}
\begin{abstract}
We argue that quantum Liouville field theory supplemented with a
suitable source term is the effective theory which describes the
short--range correlations of the gluon saturation momentum in the
two--dimensional impact--parameter space, at sufficiently high
energy and for a large number of colors. This is motivated by
recent developments concerning the stochastic aspects of the
high--energy evolution in QCD, together with the manifest scale
invariance of the respective evolution equations and general
considerations on the uncertainty principle. The source term
explicitly breaks down the conformal symmetry of the (pure)
Liouville action, thus introducing a physical mass scale in the
problem which is identified with the average saturation momentum.
We construct this source term for the case of a homogeneous
distribution and show that this leads to an interesting theory:
the relevant correlation functions are ultraviolet {\em finite}
(and not just renormalizable) when computed in perturbation
theory, due to mutual cancellations of the tadpole divergences.
Possible generalizations to inhomogeneous source terms are briefly
discussed.

\end{abstract}
\end{center}}

\end{frontmatter}

%\newpage
%\tableofcontents
\newpage

\section{Introduction}
\setcounter{equation}{0} \label{SECT_INTRO}

Recently, there has been significant progress in our understanding
of the dynamics in perturbative QCD at high energy/high gluon
density, leading to a consistent picture of the high--energy
evolution as a classical stochastic process of a special type
\cite{IMM04,IT04}: a non--local generalization of the
reaction--diffusion process of statistical physics, with the
`non--locality' referring both to the transverse momenta and to
the position in the two--dimensional `impact--parameter space' ---
the plane transverse to the collision axis. However, the
consequences of this new picture for the dynamics in
impact--parameter space are still to be explored: although the
impact--parameter dependence is in principle encoded in the
underlying evolution equations --- the Pomeron loop equations of
Refs. \cite{IT04,MSW05,IT05} ---, this dependence turns out to be
too complicated to deal with in practice, and so far it has been
neglected in the applications of these equations. Some important
aspects of this dynamics --- like the peripheral dynamics
responsible for the Froissart growth of the total cross--section
(see, e.g., \cite{FB,KW02}) ---  are clearly non--perturbative.
Still, with increasing energy, there is an increasingly large
region around the center of the hadron where the gluon density is
high and perturbation should apply, including for a calculation of
the correlations in $\bm{b}$ (the two--dimensional vector denoting
the impact parameter).

For long time, this high--density central region has been assumed
to be quasi--homogeneous --- a disk which appears uniformly black
to any external probe whose resolving power $Q^2$ in the
transverse space is smaller than some critical value fixed by the
gluon density in the target. Such a uniform disk would be
characterized by just two quantities: \texttt{(i)} the (local)
saturation momentum $Q_s^2(\bm{b})$, which fixes the critical
scale for `blackness' (i.e., for the onset of unitarity
corrections) and is roughly independent of $\bm{b}$ (it typically
has a smooth profile decreasing from the center towards the
periphery on a radial distance set by the `soft' QCD scale
$\Lambda_{\rm QCD}$), and \texttt{(ii)} the radius $R$ of this
central region where the density is high. Both quantities are
expected to grow with the energy, but whereas the respective
growth is rapid (power--like) for the saturation momentum and can
be computed in perturbation theory --- since determined by
quasi--local gluon splitting processes within the high--density
region ---, that of the `black disk' radius is much slower
(logarithmic in $s$) and also non--perturbative, since it is
related to the expansion of the black disk into the outer corona
at relatively low density.

\begin{figure}
\begin{center}
\centerline{\epsfig{file=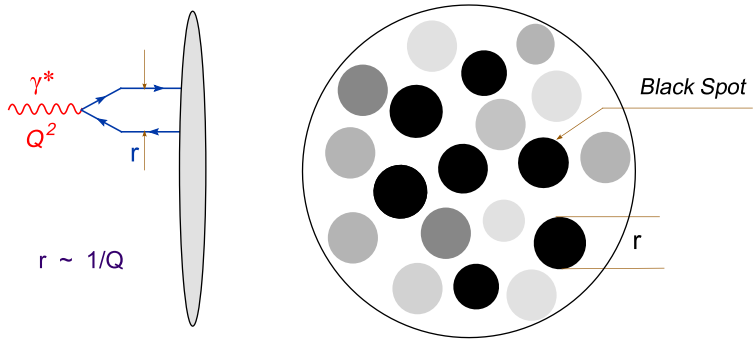,height=5.cm}}
 \caption{A transverse view of a high--energy hadron,
 as `seen' in deep inelastic scattering at very high energy. The
 dipole projectile probes a target area of the order of its own
 transverse area, at various impact parameters.}
\label{BSfig}
\end{center}
\end{figure}

Very recently, however, this traditional picture has been
challenged \cite{IT04,HIMST06,GLUON} by the newly developed
picture of the stochastic evolution, which suggests that, for
sufficiently high energies at least, the central region at high
density should be the site of {\em wild fluctuations}, leading to
the coexistence of regions where the gluon density is anomalously
high --- i.e., the local saturation momentum $Q_s^2(\bm{b})$ is
considerably larger than the respective {\em average} value $\bar
Q_s^2$ --- with regions which are anomalously dilute, such that
$Q_s^2(\bm{b})\ll \bar Q_s^2$. The high--density regions appear as
{\em black spots} to an external probe with a resolution $Q^2\sim
\bar Q_s^2$, whereas the low--density ones rather look `grey' on
the same resolution scale, with the precise nuance of `grey'
depending upon the ratio between the local saturation scale
$Q_s^2(\bm{b})$ and its average value $\bar Q_s^2$ (see  Fig.
\ref{BSfig}).
% (more precisely, upon the {\em logarithm} of this ratio; see below).
Remarkably, it turns out that, at least for a simple projectile
like a dipole, the `black spots' completely dominate the {\em
average} scattering amplitude $\langle T(Q^2)\rangle$ up to very
high resolution scales $Q^2$, well above $\bar Q_s^2$ \cite{IT04}.
That is, even for values $Q^2$ which are so high that the
scattering is weak {\em on the average}, i.e., such that $\langle
T(Q^2)\rangle\ll 1$, the average amplitude is still controlled by
the dense fluctuations for which $T(Q^2)\sim 1$, i.e., by the rare
events in which the projectile has hit a black spot.

This new picture has rich and interesting consequences: it leads
to a total breakdown of the `twist expansion' up to very high
values for $Q^2$ and it predicts a new scaling law for $\langle
T(Q^2)\rangle$ at high energies \cite{IMM04,IT04,MS04}
--- known as {\em diffusive scaling} \cite{HIMST06} ---, which should
replace the {\em geometric scaling} expected
\cite{SCALING,MT02,MP03} from mean field approximations like the
Balitsky--Kovchegov equation \cite{B,K} or the more general
(functional) JIMWLK equation \cite{JKLW,W,CGC}.

The strong fluctuations in the gluon distribution at high energy
find their origin in gluon--number fluctuations in the early
stages of the evolution, i.e., BFKL--like \cite{BFKL,AM94}
splitting processes\footnote{The `black spots' have been
originally observed \cite{AMSalam95} in numerical simulations of
Mueller's dipole picture \cite{AM94}
--- the large--$N_c$ version of the BFKL evolution ---, but at that
time it was not clear whether they would survive after including
the `unitarity corrections' (i.e., the saturation effects due to
the non--linear gluon dynamics at high density
\cite{GLR,B,JKLW,CGC}).}, which induce correlations between the
gluons having a common ancestor, and thus between the `spots'
generated via the subsequent evolution of these gluons. So far,
the effects of these correlations have been studied only at fixed
impact parameter, via a coarse--graining of the Pomeron loop
equations in impact--parameter space \cite{IMM04,IT04}. This
approximation is perhaps sufficient for a study of the scattering
of a small projectile which is quasi--localized in $\bm{b}$ (e.g.,
the case of the dipole scattering, as relevant for $\gamma^*p$
deep inelastic scattering \cite{HIMST06}, and also for $pp$ or
$pA$ collisions under specific circumstances \cite{GLUON}), but on
the other hand it prevents one from studying the {\em
correlations} in $\bm{b}$, which could be experimentally accessed
via multiple--particle production in hadron--hadron collisions
(say, at LHC). { Moreover, as we shall later argue, this
approximation has some shortcomings even for the description of
the short--range correlations, for which it was {\em a priori}
intended: by ignoring any information about the {\em size} of the
fluctuations in impact--parameter space, it artificially
suppresses the fluctuations with very small sizes, i.e., the tiny
spots where the local saturation momentum is much larger than the
average one. Or, as alluded to before, a proper description of
such tiny spots is essential in order to study the scattering of
small projectiles with resolution $Q^2\gg Q_s^2$.

In this paper, we shall make a first attempt to go beyond the
coarse--graining approximation introduced in Refs.
\cite{IMM04,IT04}. Namely, we shall propose an {\em effective
field theory} describing the distribution of the saturation
momentum in impact--parameter space at a fixed (high) energy and
for a large number of colors $N_c$. (The large--$N_c$
approximation is needed in order to be able to neglect long--range
color exchanges between saturated spots; see Sect.
\ref{SECT_Local} for details.) Note that the saturation momentum
$Q_s(\bm{b})$ is also the typical transverse momentum of the gluon
configuration located around $\bm{b}$. Hence, the effective theory
that we shall propose encodes information about the gluon
distributions in both momentum space and impact--parameter space.

Let us start by emphasizing that the very existence of an
effective field theory for $Q_s(\bm{b})$ is not at all obvious.
For instance, it is not {\em a priori} clear whether one can
effectively replace (even if only approximately) the information
about the gluon distribution or the dipole scattering amplitudes
by a theory for the distribution of the saturation momentum alone.
Furthermore, even assuming that such a theory exists, it is not at
all clear whether it can be made {\em local}. The Pomeron loop
equations are non--local in the transverse momenta and
coordinates, and this may translate into non-localities in
$\bm{b}$ in the effective theory that we are looking for.
Moreover, the correlations induced by the evolution are generally
non--local in rapidity, since associated with gluon splittings in
the intermediate steps of the evolution. Still, as we shall argue
in Sect. \ref{SECT_Local}, the {\em short--range} correlations, at
least, are quasi--local in rapidity, since predominantly produced
via gluon splittings in the late stages of the evolution, close to
the final rapidity $Y$. By `short--range' we mean distances of the
order of, or smaller than, the average saturation length
$1/Q_0(Y)$ at rapidity $Y$ (with the notation\footnote{Note that
so far we have introduced two different notations for the `average
saturation momentum', namely $\bar Q_s^2$ and $Q_0^2\equiv\langle
Q_s^2\rangle$; they correspond to different definitions, as we
shall explain in Sect. 2.} $Q_0^2\equiv\langle Q_s^2\rangle$).
Hence, we expect a local field theory to be meaningful, at least,
for the dynamics over such short distances.

To proceed, we shall simply {\em assume} that a local field theory
for $Q_s(\bm{b})$ exists and then we shall constrain its structure
from general physical considerations. More precisely, we shall
construct this theory as the natural generalization of the
$\bm{b}$--independent distribution proposed in Refs.
\cite{IMM04,IT04} which is consistent the {\em uncertainty
principle} and the {\em conformal symmetry} of the high--energy
evolution equations. Together, these constrains almost uniquely
fix the structure of the effective theory, as we explain now: }

As already mentioned, the saturation momentum $Q_s(\bm{b})$ is
also the typical transverse momentum of the gluon configuration
centered at $\bm{b}$. By virtue of the uncertainty principle, the
area occupied by that configuration cannot be {\em smaller} than
$1/Q_s^2(\bm{b})$. This area cannot be much {\em larger} either,
since a gluon configuration which has reached saturation on some
scale $Q_s(\bm{b})$ can hardly emit gluons with soft momenta
$k_\perp\ll Q_s(\bm{b})$ \cite{JKMW97,AM99,SAT,GAUSS}. That is,
its subsequent evolution predominantly proceeds via the emission
of harder gluons with $k_\perp\gg Q_s(\bm{b})$, which cannot
increase the size of the configuration (except very slowly, via
peripheral emissions). In turn, such hard gluons become spots
which evolve towards saturation on their own sizes and at the same
time act as sources for even smaller spots, which spread over the
surface of the original configuration and where the local
saturation momenta are much harder. We see that the high--energy
evolution is strongly biased towards smaller sizes, thus leading
--- in a three--dimensional picture where the impact--parameter
space occupies the $(x,y)$--plane and the saturation momentum
$Q_s(\bm{b})$ is represented along the $z$--axis --- to a
`landscape' picture, with spikes of various heights, randomly
distributed and surrounded by valleys.

Furthermore, the equations describing the evolution in QCD at high
energy and in the leading logarithmic approximation have the
important property of {\em conformal symmetry}, that is, they are
invariant under conformal (M\"obius) transformations in the
transverse plane. This has since long been appreciated in the
context of the linear, BFKL, evolution \cite{BFKL}, with important
consequences \cite{lipatov,lipatov1,FK95} (in particular, for the
calculability of the theory; see Ref. \cite{ewerz} for a review
and more references), but this remains true for the non--linear BK
\cite{B,K}, JIMWLK \cite{JKLW,W,CGC}, or Pomeron loop
\cite{IT04,MSW05,IT05}, equations, since all these equations
involve the BFKL splitting kernel (for gluons or dipoles) together
with gluon--number changing vertices, which indeed respect
conformal symmetry. One may think that this symmetry is
inconsistent with the emergence of a special scale --- the
(average) saturation momentum
--- in the solutions to these equations, but this is actually not
true: the saturation scale is not spontaneously generated by the
evolution, rather it comes out as the evolution of the scale
introduced by the initial conditions at low energy
\cite{GLR,AM99,SCALING,MT02}. On the other hand, the gluon {\em
correlations} are generated by the conformally--invariant
evolution\footnote{We assume that there were no correlations in
the initial conditions at low energy, for simplicity.}, so we
expect this symmetry to constrain the effective theory describing
these correlations.

The effective theory that we shall arrive at by exploiting such
considerations is a two--dimensional, interacting, scalar field
theory, in which the field $\phi$ is proportional to the logarithm
of the local saturation momentum, and which involves two `free'
parameters: the expectation value $Q_0^2\equiv\langle
Q_s^2\rangle$ of the saturation momentum and a dimensionless
coupling constant $\sigma$, which characterizes the degree of
disorder introduced by gluon--number fluctuations in the course of
the evolution. Both parameters are expected to rise with the
energy, in a way which is in principle determined by the
underlying QCD evolution equations, but which needs not be
specified for our present purposes. It suffices to say that the
weak coupling regime $\sigma\ll 1$ in the effective theory
corresponds to low, or intermediate, energies, where the dynamics
is quasi--deterministic, whereas the strong coupling regime
$\sigma\gg 1$ corresponds to the more interesting situation at
high energy, where the stochastic aspects are fully developed and
essential.

The effective theory involves three basic ingredients:
\texttt{(i)} the standard kinetic term, which couples fluctuations
at neighboring points, \texttt{(ii)} a potential term, which
ensures that the typical gradients at $\bm{b}$ are of order
$Q_s^2(\bm{b})$, in fulfillment of the uncertainty principle, and
\texttt{(iii)} a source term, which enforces the value
$Q_0^2\equiv\langle Q_s^2(\bm{b})\rangle$ of the average
saturation momentum. Since proportional to
$Q_s^2\propto\rme^{\sigma\phi}$, the potential is exponential in
$\phi$, which is precisely as it should for consistency with
conformal symmetry: together, the kinetic plus the potential terms
are recognized as the {\em Liouville action}, which is
conformally--invariant, and thus integrable. The quantum Liouville
field theory (LFT) has been extensively studied over the last
decades --- especially, in connection with studies of quantum
gravity in two dimensions --- and many exact results are known by
now about its properties
\cite{Poly81,Se90,GinMoore,DO94,ZZ96,Tesch}. However, precisely by
virtue of its symmetry, LFT involves no mass scale (it gives rise
to power--law correlations on all distance scales), and thus
cannot accommodate a non--zero value for the average saturation
momentum. This is taken care off by the source term, linear in
$\phi$, at the expense of explicitly breaking the conformal
symmetry. This breaking has dramatic consequences on the
properties of the theory and it also complicates its analysis very
much (since it is not possible to directly exploit the large
amount of information known about LFT).

Yet, some general properties of the effective theory can be
inferred by semi--classical and perturbative techniques, or simply
by inspection of the action. The presence of the source term
endows this theory with a stable ground state (in contrast to
LFT), which appears as a saddle point of the action and allows for
a perturbative treatment of the weak coupling ($\sigma\ll 1$)
regime. The perturbative calculations, that we shall push up to
two--loop order, reveal a very interesting property, which is
furthermore confirmed, at non--perturbative level, by inspection
of the corresponding Dyson equations: the correlations of the
saturation momentum (an exponential operator in this theory) are
{\em ultraviolet finite}, and not just renormalizable, as
generally expected for a two--dimensional field theory. {This
means that the operator $Q_s^2$ has no anomalous dimension, which
in turn implies that its short--range correlations, over distances
$R\simle 1/Q_0$, have a {\em power--like} decay, with exactly the
{\em same} powers as in pure Liouville theory --- since these
powers are fixed by the natural dimension of the operator. On the
other hand, the theory predicts that these correlations decay {\em
exponentially} on larger distances $R \gg 1/Q_0$, with the
(average) saturation momentum $Q_0$ playing the role of a
screening mass. The emergence of an exponential fall--off in the
context of perturbative QCD may look surprising, but it is
presumably an artifact of our insistence on a field theory which
is local in rapidity: the physical correlations on large distances
$R\gg 1/Q_0$ are typically generated via gluon splittings in the
early stages of the evolution (the earlier, the larger $R$ is; see
Sect. \ref{SECT_Local})), which are not encoded in our present
theory.}

The paper is organized as follows: Sect. \ref{SECT_Local} presents
a brief and critical discussion of the $\bm{b}$--independent
distribution proposed in Refs. \cite{IMM04,IT04}, with the purpose
of clarifying its limitations and, more generally, the limitations
of any effective theory which is local in rapidity. In Sect.
\ref{SECT_Weak} we construct the low--energy/weak coupling
($\sigma\ll 1$) version of the effective field theory, as the
straightforward extension of the $\bm{b}$--independent
distribution alluded to above. Sect. \ref{SECT_Strong} is our main
section: after a quick introduction to the Liouville field theory
and its conformal symmetry, we explain the relevance of this
theory for the QCD problem at hand, then present the complete
action for our effective theory (including the symmetry--breaking
source term) and discuss some of its properties. Sect.
\ref{SECT_UV} is slightly more technical, since devoted to
perturbative calculations to two--loop order, with the purpose of
demonstrating the ultraviolet--finiteness of the correlation
functions when computed in the effective theory. Finally, Sect. 6
summarizes our results and conclusions. In the Appendix, we show
how to couple the effective theory to the CGC formalism (within
the simple context of McLerran--Venugopalan model \cite{MV}),
which is useful in view of computing observables.

\section{The coarse--graining approximation and its limitations}
\setcounter{equation}{0}\label{SECT_Local}

Our starting point is the Gaussian probability distribution for
the saturation momentum introduced in Refs. \cite{IMM04,IT04} (see
also Refs. \cite{BD97,BDMM,MSX06}), which can be viewed as a
coarse--grained version of the effective theory that we intend to
construct, in which the impact--parameter dependence has been
averaged out. The applicability of such a coarse--graining will be
shortly discussed.

The random variable in this distribution is the logarithm
$\rho_s\equiv \ln(Q_s^2/\Lambda^2)$ of the saturation momentum
(with $\Lambda$ some arbitrary scale of reference), which is the
scale which separates, in an event--by--event description, between
a high--density, or `Color Glass Condensate' \cite{EDICGC}, phase
at low transverse momenta ($k_\perp < Q_s$ or $\rho\equiv
\ln(k_\perp^2/\Lambda^2)<\rho_s$), where the gluon occupation
factor saturates\footnote{Strictly speaking, the gluon occupancy
keeps growing with the energy in the CGC phase, but only very
slowly: logarithmically in $s$, that is, linearly in the rapidity
$Y=\ln s$ \cite{AM99,SAT}.} at a large value of
$\mathcal{O}(1/\alpha_s)$, and a low density phase at high momenta
($\rho > \rho_s$), where the gluon occupancy is still low but it
rises rapidly with the energy (as a power of $s$), via BFKL gluon
splitting. It is customary to work with the `rapidity' variable
$Y\equiv \ln s$, which plays the role of an `evolution time' for
the high--energy evolution. When increasing $Y$, modes with higher
and higher values of $k_\perp$ enter at saturation, so the
borderline $\rho_s$ propagates towards higher transverse momenta.
However, due to gluon--number fluctuations in the splitting
process --- which can be associated with the fact that the
particle number is discrete \cite{IMM04} ---, this progression of
$\rho_s$ is not uniform but stochastic, like a one--dimensional
Brownian motion.

Based on the analogy with the reaction--diffusion problem
\cite{BD97,SaarPanja,BDMM,MSX06}, it has been argued in Refs.
\cite{IMM04,IT04} that $\rho_s$ should be a Gaussian random
variable with probability distribution
\begin{equation}\label{eq-prob}
    P_Y(\rho_s) =
    \frac{1}{\sqrt{2\pi}\sigma}\,
    \exp \left[
    -\frac{\left( \rho_s - \langle \rho_s \rangle \right)^2}{2\sigma^2}
    \right]\,,
\end{equation}
where both the central value $\langle \rho_s \rangle$ and the
dispersion $\sigma^2=\lan\rho_s^2\ran - \langle\rho_s\rangle^2$
are expected\footnote{Recall that, throughout this analysis, we
consider the leading--order approximation where the coupling is
not running.} to rise linearly with $Y$ (for $Y$ large enough):
$\langle \rho_s \rangle=\lambda\bar\alpha_s Y$ and $\sigma^2
\simeq D\bar\alpha_s Y$, with $\bar\alpha_s\! =\!\alpha_sN_c/\pi$.
The coefficients $\lambda$ and $D$ are known \cite{BD97,BDMM} only
in the formal limit $\alpha_s\to 0$, and the corresponding
expressions cannot be extrapolated to realistic values of
$\alpha_s$ since they depend upon $\ln(1/\alpha_s^2)$. For what
follows, their actual values are unimportant, and so are the
precise $Y$--dependencies of $\langle \rho_s \rangle$ and
$\sigma$; all that matters is that these quantities rise quite
fast with $Y$. So long as $\sigma\ll 1$ --- the case for
relatively low energies --- the theory remains
quasi--deterministic. But the most interesting regime for us here
is the {\em high--energy regime} at $\sigma \ge 1$, where the
fluctuations are fully developed and have a strong influence on
the theory.

\begin{figure}
\begin{center}
\centerline{\epsfig{file=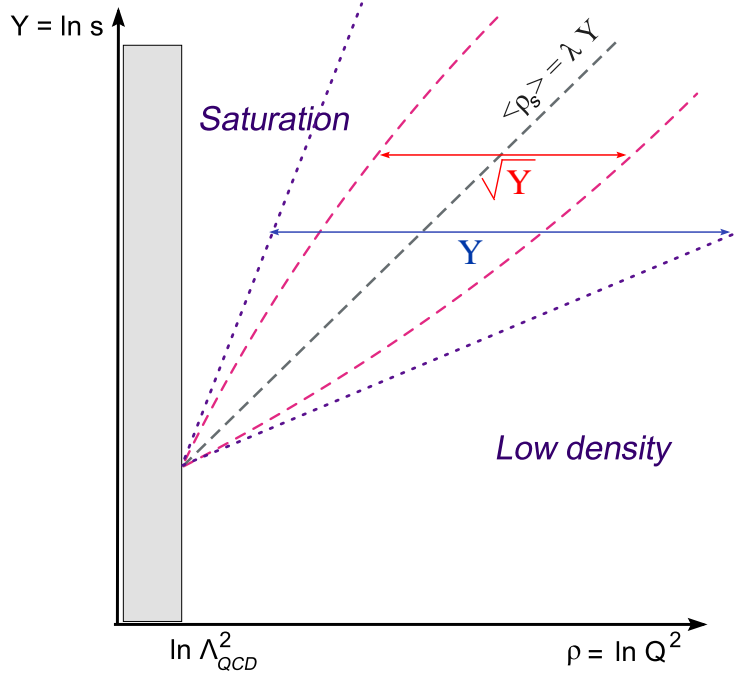,height=9.cm}}
 \caption{A picture of the diffusive saturation boundary in the
 presence of fluctuations; shown are the average saturation line
 $\langle \rho_s \rangle=\lambda\bar\alpha_s Y$,
 the diffusive radius $\sigma\propto\sqrt{Y}$
 for the Gaussian distribution (\ref{eq-prob}), and the
 diffusive scaling window with a width $\sigma^2\propto Y$.}
\label{DSWfig}
\end{center}
\end{figure}

Since the saturation momentum is fluctuating, a physical
discussion will naturally involve a {\em range} of values for
$Q_s$, rather than just a single value (see Fig. \ref{DSWfig}).
The range which is {\em a priori} privileged by the Gaussian
probability (\ref{eq-prob}) is the `diffusive disk' at $|\rho_s
-\langle \rho_s\rangle| \le \sigma$, where the probability is
reasonably large: $P(\rho_s)\simge 1/\sigma$. But the relevant
range also depends upon the physical quantity of interest, which
can favor some values of $Q_s$ over the others. The simplest
example in that sense refers to the very definition of the
`average saturation momentum', which is not unique (because the
relation between $Q_s$ and $\rho_s$ is non--linear: $Q_s^2=
\Lambda^2 {\rm e}^{\rho_s}$) and thus needs to be properly
specified --- different definitions can be relevant for different
problems. The form of the Gaussian distribution (\ref{eq-prob})
makes it natural to define
 \beq\label{eq-QSdef1}
 \bar Q_s^2\,\equiv\,\Lambda^2 \,{\rm e}^{\langle \rho_s
 \rangle}\,.
 %\,=\,\Lambda^2 {\rm e}^{\lambda \abar Y}\,,
 \eeq
But the expectation value of the `operator' ${\rm e}^{\rho_s}$ can
be also computed, with the following result:
  \beq\label{eq-QSdef2}
  \langle Q_s^2 \rangle\,\equiv\,\Lambda^2 \,\langle
 {\rm e}^{\rho_s}\rangle\,=\,\Lambda^2\,{\rm e}^{\langle \rho_s
 \rangle}\, {\rm e}^{\frac{1}{2}\sigma^2}\,=\,
 \bar Q_s^2 \, \,{\rm e}^{\frac{1}{2}\sigma^2}
 % \Lambda^2 \,{\rm e}^{(\lambda +D/2)\abar Y }
 \,.
 \eeq
This is always harder than the scale in Eq.~(\ref{eq-QSdef1}), and
for sufficiently high energies, where $\sigma \gg 1$, it is even
{\em much} harder. This considerable difference between the two
definitions can be easily traced back: in computing $\langle Q_s^2
\rangle$, the exponential operator ${\rm e}^{\rho_s}$ biases the
integration towards very large values of $\rho_s$, of order
$\rho_s\sim \langle \rho_s\rangle + \sigma^2$, as opposed to the
typical value $\rho_s\sim \langle \rho_s\rangle$ contributing to
$\bar Q_s^2$. Note that the large deviation $\rho_s -\langle
\rho_s\rangle \sim \sigma^2$ corresponds to fluctuations which are
very rare: $P(\rho_s)\sim \exp(-\sigma^2/2)\ll 1$, and which
occupy a very small area $\sim 1/\langle Q_s^2 \rangle$ in
impact--parameter space (by the uncertainty principle). Yet, such
rare and tiny fluctuations are physically important, as they
define the upper bound of the {\em diffusive scaling window}
\cite{IT04,HIMST06,GLUON}:
   \beq\label{eq-diffscaling}
 - \frac{\sigma^2}{2}\,\ll\,  \rho -\langle \rho_s\rangle \,\ll\,
 \frac{\sigma^2}{2}\,,
 \eeq
(see also Fig. \ref{DSWfig}) where $\rho\equiv
\ln(Q^2/\Lambda^2)$, with $Q^2$ the transverse resolution of a
small projectile, like a `color dipole' (a $q\bar q$--pair
fluctuation of the virtual photon in deep inelastic
electron--hadron scattering), which scatters off the hadron. The
interval in Eq.~(\ref{eq-diffscaling}) represents the kinematical
window at high energy ($\sigma \gg 1$) within which the (average)
dipole scattering amplitude $\langle T(\rho,Y)\rangle$  `scales'
as a function of the dimensionless variable $Z\equiv (\rho
-\langle \rho_s\rangle)/\sigma$ rather than separately depending
upon $\rho$ and $Y$ : $\langle T(\rho,Y)\rangle\approx \langle
T(Z)\rangle$ (see Refs. \cite{IT04,HIMST06,GLUON} for details).
The diffusive scaling window in Eq.~(\ref{eq-diffscaling}) can be
alternatively described as
  \beq\label{eq-diffscaling1}
 \bar Q_s^2 \,{\rm e}^{-\frac{1}{2}\sigma^2}
 \,\ll\,\,  Q^2 \,\ll\, \,
 \bar Q_s^2 \,{\rm e}^{\frac{1}{2}\sigma^2}
 \equiv\, \langle Q_s^2 \rangle\,,
 \eeq
which explicitly shows that both definitions for the `average
saturation momentum' introduced in Eqs.~(\ref{eq-QSdef1}) and,
respectively, (\ref{eq-QSdef2}) play a role in characterizing the
physical effects of the fluctuations in $Q_s$.

At this point one should remember that the previous discussion was
based on the local approximation (\ref{eq-prob}) for the
distribution of $Q_s$, so it is important to understand what are
the assumptions beyond this approximation. In Refs.
\cite{IMM04,IT04} (see, especially, the discussion in Sect. 6 of
Ref. \cite{IT04}), the distribution (\ref{eq-prob}) has been
obtained  via a {\em coarse--graining} in impact--parameter space:
the $\bm{b}$--dependence of the dipole scattering amplitudes, as
encoded in the Pomeron loop equations \cite{IT05,MSW05}, has been
averaged out over a region in the impact--parameter space which in
Refs. \cite{IMM04,IT04} has been loosely characterized as the
`dipole size', but which should be more properly interpreted as an
intrinsic scale in the hadron, so like $1/\bar Q_s$. In this
averaging, one has potentially neglected two types of
correlations: \texttt{(i)} {\em short--range} correlations between
the fluctuations (`spots') with small sizes $R\ll 1/\bar Q_s$
(i.e., with large saturation momenta $Q_s\gg \bar Q_s$) which lie
inside the coarse--graining cell --- such smaller spots were
treated as being uniformly distributed over the cell area $1/\bar
Q_s^2$, and \texttt{(ii)} the {\em long--range} ($R\gg 1/\bar
Q_s$) correlations between different cells. Besides, one has
assumed {\em homogeneity} in $\bm{b}$ --- the average saturation
momentum $\bar Q_{s}^2$ and the dispersion $\sigma^2$ were taken
to be independent of $\bm{b}$
---, which is not really an approximation, but merely a choice for
the initial conditions at low energy (the homogeneity being
preserved by the evolution according to the Pomeron loop
equations). Under these assumptions, the $\bm{b}$--dependence has
disappeared from the evolution equations, which were then shown
\cite{IT04} to be equivalent to a stochastic equation of the sFKPP
type --- the Langevin equation for the reaction--diffusion process
\cite{SaarPanja}. The Gaussian probability distribution in
Eq.~(\ref{eq-prob}) then follows from known properties of the
sFKPP equation, as recently clarified in the respective literature
\cite{BD97,BDMM,MSX06}.

Thus, the fact that here is no explicit $\bm{b}$--dependence in
Eq.~(\ref{eq-prob}) must be understood as follows: this formula is
meant to apply to points $\bm{b}$ within a given coarse--graining
cell, with an area $\sim 1/\bar Q_{s}^2\,$; at all such points,
the event--by--even local saturation scale is assumed to be the
same, and equal to $\rho_s$ (in logarithmic units); that is, all
the points within a cell fluctuate {\em coherently} with each
other, so like a rigid body. On the other, nothing is said about
the correlations between different cells: Eq.~(\ref{eq-prob})
apply only to {\em local} fluctuations within a given cell, as
averaged over the size of that cell.

{How does this approximate picture compare with the physical
reality ? There are {\em a priori} two mechanisms for building
spot--spot correlations: \texttt{(a)} {\em Long--range
interactions between saturated spots:} A gluon configuration with
saturation momentum $Q_s$ has a low but non--zero probability to
emit low--momentum gluons with $k_\perp\ll Q_s$, thus creating
long--range color fields over distances $R\gg 1/Q_s$. These are
dipolar fields, and not Coulomb ones, because the total color
charge gets screened at saturation \cite{SAT,AM02,GAUSS}. These
fields mediate dipole--dipole interactions between saturated spots
which are far away from each other. However, these interactions
are suppressed by a factor $1/N_c^2$, as they imply color
exchanges, and thus can be safely neglected in the large--$N_c$
approximation underlying the present discussion. \texttt{(b)} {\em
Gluon--number fluctuations in the high--energy evolution:} As
mentioned in the Introduction, different spots can be correlated
with each other because they have a common ancestor at some
earlier rapidity. Such correlations survive at large $N_c$, since
associated with fluctuations in a colorless quantity: the number
of gluons in the light cone gauge. Eq.~(\ref{eq-prob}) is meant to
capture these fluctuations in the coarse--graining approximation,
and our purpose here is to relax this approximation by restoring
the $\bm{b}$--dependence of the induced correlations.

However, as also mentioned in the Introduction, the relevant
correlations are generally {\em non--local in rapidity}, since
induced through branching processes which can occur at all the
intermediate rapidities. As we shall explain now\footnote{We would
like to thank Al Mueller for illuminating discussions on this
particular issue.}, the {\em short--range} correlations generated
in this way can nevertheless be encoded in a {\em local} (in $Y$)
field theory, of the type we would like to construct. To that aim,
we shall consider two spots which are separated by a distance $R$
at the final rapidity $Y$. These spots are correlated with each
other provided they had a common ancestor at some earlier rapidity
$0<y<Y$. It turns out that $y$ is correlated with $R$. Indeed, the
typical spots at rapidity $y$ have a saturation momentum $\bar
Q_{s}(y)$ and a size $\sim 1/\bar Q_{s}(y)$. If that size is much
smaller than $R$, then such a typical spot has only little
probability to emit a gluon at a distance $R$ away from its center
and thus initiate the evolutions leading to the two final spots
that we measure at $Y$. (This probability decreases as a large
power of $(R\bar Q_{s}(y))^{-1}$ because of the screening of the
color charge at saturation.) If, on the other hand, $1/\bar
Q_{s}(y)\gg R$, then there is a geometrical penalty factor $\sim
(R\bar Q_{s}(y))^2$ for the gluon to be emitted inside the
(relatively small) domain with area $R^2$. Hence, the most
important intermediate rapidity $y$ for creating correlations over
a distance $R$ is the one for which $\bar Q_{s}(y)\sim 1/R$.
Therefore, the larger is $R$, the more we have to go backwards in
the evolution, and the more non--local are the respective
correlations in $Y$. Vice--versa, the short--range correlations
with $R\simle 1/\bar Q_{s}(Y)$ are typically produced in the late
stages of the evolution, at $y\sim Y$, and hence are quasi--local
in rapidity. It should be therefore possible to encode such
short--range correlations in a local, effective, field theory.

As we shall later see, these considerations are indeed consistent
with the effective field theory that we shall arrive at, except
for the fact that the actual scale for correlations which will
emerge from that theory is not the scale $\bar Q_s^2$ that we have
focused on in the above discussion, but rather the harder scale
$\langle Q_s^2 \rangle$, cf. Eq.~(\ref{eq-QSdef2}). }

%The reason for this difference is because a quantum field theory
%favors the high--momentum fluctuations, which have small sizes and
%hence are costless in action.

\section{The effective field theory: weak coupling}
\setcounter{equation}{0}\label{SECT_Weak}

With this section, we start our program aiming at extending the
coarse--grained distribution in Eq.~(\ref{eq-prob}) to an
effective field theory which describes the distribution of the
saturation momentum in impact--parameter space. For more clarity,
we shall first develop our arguments for the low energy/weak
coupling regime $\sigma\ll 1$, where we shall argue that the
corresponding extension is a free field theory for a massive
scalar field in two (Euclidean) dimensions. Then, in Sect.
\ref{SECT_Strong}, we shall present the generalization of this
theory to arbitrary values of the coupling $\sigma$.

For more clarity, we shall assume `mean-field--like' initial
conditions at low energy ($Y=0$): the initial gluon density is
large and homogeneous, and the saturation momentum $Q_s(Y=0)$
takes the same value $\Lambda$ (with $\Lambda^2\gg \Lambda^2_{\rm
QCD}$) at all the points. By boosting this system to high energy,
the gluon distribution evolves via (generally, non--linear) gluon
splitting, and inhomogeneities (`spots') appear in the
event--by--event description, due to gluon--number fluctuations.

In the early stages of this evolution --- namely, so long as
$\sigma\ll 1$ ---, the fluctuations have no time to significantly
develop, so all the spots have more or less the same
size\footnote{Note that, when $\sigma\ll 1$, the various `average
saturation scales' characterizing the statistical ensemble (cf.
Sect. \ref{SECT_Local}) are close to each other: $\bar Q_s^2
\approx \langle Q_s^2 \rangle$.}, $\sim 1/\bar Q_s$, and the same
value for the saturation momentum, $\sim \bar Q_s$; furthermore,
they are only weakly correlated with each other. It is then
straightforward to extend the coarse--grained approximation
(\ref{eq-prob}) to a distribution which covers the whole
transverse profile of the hadron: this is simply the product of
independent Gaussian distributions like that in
Eq.~(\ref{eq-prob}), one for each spot :
 \begin{equation}\label{eq-probi}
    P_Y[\rho_s] = \prod_j
    \frac{1}{\sqrt{2\pi}\sigma}\,
    \exp \left[
    -\frac{\left( \rho_s(j) - \langle \rho_s \rangle \right)^2}{2\sigma^2}
    \right]\quad\mbox{for}
 \quad \sigma\ll 1\,,
\end{equation}
where the discrete variable $j$ labels the spots, and each spot
has roughly an area $1/\bar Q_s^2$. Introducing the fluctuation
field $\eta(j)\equiv \rho_s(j) - \langle \rho_s \rangle$,
Eq.~(\ref{eq-probi}) implies: $\langle \eta(j)\eta(l)\rangle=
\sigma^2\delta_{lj}$. So far, the quantities $\rho_s(j)$ or
$\eta(j)$ refer globally to a spot. In order to promote them to
{\em field} variables $\rho_s(\bm{x})$ or $\eta(\bm{x})$, with
$\bm{x}$ denoting the impact parameter, but keep the same
correlations as above, one needs to introduce a kinetic term to
smear out correlations over the spot size $\sim 1/\bar Q_s$. The
simplest kinetic term, and also the only one to be renormalizable
in two dimensions, is the standard, quadratic, kinetic term, that
we shall adopt in what follows. We are thus led to the following
field--theoretical extension of Eq.~(\ref{eq-probi})
  \beq\label{eq-S0}
  P_Y[\eta]\,=\,\exp\left\{-S_0[\eta]\right\},\quad\mbox{with}
 \quad S_0[\eta]\equiv \,\frac{1}{2\sigma^2}\int \dif^2 \bm{x}
 \left\{ (\grad^i \eta)^2 + \bar Q_s^2 \eta^2\right\}\,,\eeq
which provides the weight function for the functional integral
over $\eta(\bm{x})$ which defines expectation values in the
framework of this effective field theory. The measure in the
functional integral is assumed to carry the proper normalization:
$\int D[\eta]\,\exp\left\{-S_0[\eta]\right\} = 1$.

Eq.~(\ref{eq-S0}) implies $ \langle
\eta(\bm{x})\eta(\bm{y})\rangle= \sigma^2\,D(\bm{x}-\bm{y})$, with
$D$ the propagator of a free, massive, scalar field in a
two--dimensional Euclidean space:
 \beq\label{eq-D0}
 D(\bm{x-y})=\int\frac{\dif^2 \bm{k}}{(2\pi)^2}
 \,\frac{\rme^{i\bm{k}\cdot \bm{(x-y)}}}{k^2+\bar Q_s^2}\,
 =\, \frac{1}{2\pi}\,{\rm K}_0(r\bar Q_s)
 \eeq
where $r=|\bm{x-y}|$ and ${\rm K}_0$ is the respective Bessel
function, with the following limiting behaviours at short and,
respective, large distances:
 \beq\label{eq-K0}
  {\rm K}_0(z)\approx\,
  \begin{cases}
         \displaystyle{\ln\frac{1}{z}\,,} &
        \text{ for\,  $z \ll 1$}
        \\*[0.5cm]
        \displaystyle{\sqrt{\frac{\pi}{2z}}\rme^{-z}\,,} &
        \text{ for\,  $z\gg 1$}.
    \end{cases}
  \eeq
We see that the correlations generated by the field theory in
Eq.~(\ref{eq-S0}) have indeed the sought for structure: they are
quasi--uniform on relatively short distances, $r\simle 1/\bar
Q_s$, (i.e., among points located within a same spot), but they
decay very fast (exponentially) over distances much larger than
the typical size of a spot.

The emergence of {\em exponentially} decaying correlations at
large distances should be taken with a grain of salt (cf. the
discussion in the Introduction): It corresponds to the fact that
the correlations over larger distances $r\gg 1/\bar Q_s$ are
predominantly produced via splittings occuring in the early stages
of the evolutions, i.e., at rapidities $y < Y$, whose effects are
not included in the present formalism, local in $Y$. As for the
late splittings (those taking place at $y\sim Y$), they produce
correlations which fall off according to power laws at large
separations $r\gg 1/\bar Q_s$, because the emissions of soft
gluons with $k_\perp\ll \bar Q_s(Y)$ is strongly suppressed by
saturation. The fact that, in our formalism, this fall--off
appears to be exponential, rather than power--like, is because we
have not provided a faithful description of the gluon spectrum,
including its softening at momenta below $Q_s$, but rather we have
replaced this whole spectrum by a unique scale --- the saturation
momentum ---, which represents the average transverse momentum of
the saturated gluons. To summarize this argument, the theory with
action (\ref{eq-S0}) provides the correct, logarithmic, behaviour
at short distances, and it mimics the rapid fall--off at larger
distances by an exponential tail, rather than the correct,
power--law, one. This will be a generic feature of the effective
theory, including at strong coupling.

We now return to the functional distribution in Eq.~(\ref{eq-S0})
and consider its predictions for the correlation functions of the
operator saturation momentum, $Q_s^2(\bm{x})\equiv \Lambda^2 {\rm
e}^{\rho_s}= \bar Q_s^2 {\rm e}^{\eta}$. One finds
  \beq\label{eq-QSweak0}
  \langle Q_s^2 (\bm{x})\rangle&\,=\,&
 \bar Q_s^2 \, \,{\rm e}^{\frac{1}{2}\sigma^2D(0)}\,,\nn
  \langle Q_s^2 (\bm{x})Q_s^2 (\bm{y})\rangle&\,=\,&
 \bar Q_s^4 \, \,{\rm e}^{\sigma^2[D(0)+D(\bm{x}-\bm{y})]}
 \,,
 \eeq
etc. Strictly speaking, these expressions should be trusted only
up to terms of $\mcal{O}(\sigma^2)$, since we are working under
the assumption that $\sigma\ll 1$; notwithstanding, we here
display the full exponentials, for comparison with subsequent
results at strong coupling.

The above formul\ae{} involve the equal--point limit $D(0)$ of the
propagator, which is divergent: $D(0)=({1}/{2\pi})\ln({1}/{a\bar
Q_s})$, with $a$ a short--distance cutoff (`lattice spacing')
introduced to regularize the divergence. We see that, even within
this free field theory, the exponential (or `vertex') operator
$V(\bm{x})\equiv{\exp}\{\eta(\bm{x})\}$ develops ultraviolet
divergences, which arise from contractions internal to this
operator. As is well known \cite{ZJ}, and also manifest on
Eq.~(\ref{eq-QSweak0}), such divergences can be eliminated via
multiplicative renormalization of the vertex operator (which here
is tantamount to normal--ordering the polynomial operators which
appear in its expansion). Namely, we shall define the renormalized
vertex operator as
 \beq\label{eq-Vren}
 V_R(\bm{x})\equiv V(\bm{x})\,{\exp}\left\{\frac{\sigma^2}{4\pi}
 \ln\left(a\rme^{2\pi}\bar Q_s\right)\right\}\,=\,
  {\exp}\left\{\frac{\sigma^2}{2}[1-D(0)]\right\}\,
 {\rm e}^{\eta(\bm{x})}\,,
 \eeq
where the particular subtraction point $1/a=\rme^{2\pi}\bar Q_s$
has been chosen for reasons to shortly become clear. One then
obtains the following, finite, vertex correlation functions:
 \beq\label{eq-Vweak}
  \langle V_R (\bm{x})\rangle\,=\,
  {\rm e}^{\frac{1}{2}\sigma^2},\ \dots \ , \
  \langle V_R (\bm{x}_1) \cdots V_R (\bm{x}_n)\rangle
  \,=\,{\exp}\Big\{\frac{n}{2}\sigma^2  + \sigma^2\sum_{i<j}
  D(\bm{x}_i-\bm{x}_j)\Big\}
 \,,
 \eeq
which in turn imply, for the renormalized operator
$Q_s^2(\bm{x})\equiv \bar Q_s^2\,V_R(\bm{x})$,
   \beq\label{eq-QSweakR}
  \langle Q_s^2 (\bm{x})\rangle&\,=\,&
 \bar Q_s^2 \, \,{\rm e}^{\frac{1}{2}\sigma^2}\,,\nn
  \langle Q_s^2 (\bm{x})Q_s^2 (\bm{y})\rangle&\,=\,&
 \bar Q_s^4 \, \,{\rm e}^{\sigma^2+ \sigma^2
 D(\bm{x}-\bm{y})}\,=\,\langle Q_s^2\rangle^2
  \,{\rm exp}\{\sigma^2 D(\bm{x}-\bm{y})\}
 \,.
 \eeq
One can now appreciate the particular choice for the subtraction
point in Eq.~(\ref{eq-Vren}): this is such that the expectation
value $\langle Q_s^2\rangle$ computed in the present field theory
precisely matches the corresponding prediction of the
coarse--grained approximation, cf. Eq.~(\ref{eq-QSdef2}).

Interestingly, Eq.~(\ref{eq-QSweakR}) reveals the emergence of
{\em power--like} correlations over (relatively) short distances,
that is, in between the points lying inside a same spot:
 \beq\label{eq-QSweakpower}
  \langle Q_s^2 (\bm{x})Q_s^2 (\bm{y})\rangle\,\approx\,
  \frac{  \langle Q_s^2\rangle^2}
  {\big[|\bm{x-y}|\bar Q_s\big]^{{\sigma^2}/{2\pi}}}
   \qquad \mbox{for}\qquad |\bm{x-y}| \ll 1/\bar Q_s\,.
 \eeq
This exhibits a singularity in the equal--point limit, which is
however physical, in the sense that one cannot localize
fluctuations in a quantum field theory down to a point without
generating singularities. On the other hand, at large distances,
the {\em connected} piece of the 2--point function in
Eq.~(\ref{eq-QSweakR}) dies away exponentially:
 \beq\label{eq-QSweakexp}
  \langle Q_s^2 (\bm{x})Q_s^2 (\bm{y})\rangle-
  \langle Q_s^2\rangle^2\,\propto\,\,\sigma^2\langle Q_s^2\rangle^2
  \exp(-|\bm{x-y}|\bar Q_s)
   \quad\ \mbox{for}\quad\ |\bm{x-y}| \gg 1/\bar Q_s\,.
 \eeq
Once again, this exponential decay should be taken with a grain of
salt: the effective theory is not supposed to apply to such large
separations.

\section{The effective field theory: general case}
\setcounter{equation}{0}\label{SECT_Strong}

Although this has not been explicitly spelled out in the previous
discussion, the free field theory of Eq.~(\ref{eq-S0}) is indeed
consistent with the uncertainty principle in the regime where this
theory is meant to apply (i.e., for $\sigma\ll 1$): indeed, the
kinetic term there smears out inhomogeneities over the very short
distances $r \ll 1/\bar Q_s$, in agreement with the fact that the
size of a spot cannot be smaller than the inverse of the typical
momentum of the gluons composing that spot, namely $k_\perp\sim
\bar Q_s$. However, for sufficiently high energy, such that
$\sigma\simge 1$, the fluctuations become important and then the
actual saturation momentum $Q_s^2 (\bm{x})$ at a given point and
in a given event can be very different from any of its expectation
values ($\bar Q_s^2$ or $\langle Q_s^2\rangle$) previously
introduced. In such a case, the uncertainty principle requires
that the minimal size of a spot located at $\bm{x}$ be fixed by
the actual value $Q_s^2(\bm{x})$ of the saturation momentum, and
not by its expectation value. This means that, in the general
action for $\eta$, the kinetic term should compete with the {\em
actual} (event--by--event) saturation momentum, and not with its
`expectation value' (whatever the meaning of the latter is).

Accordingly, the effective `mass term' in the action should be the
field--dependent scale $Q_s^2(\bm{x})\equiv \bar
Q_s^2\,{\exp}\{\eta(\bm{x})\}$. This argument suggests the
following generalization of Eq.~(\ref{eq-S0}):
 \beq\label{eq-S1}
 S_1[\eta]\equiv \,\frac{1}{2\sigma^2}\int \dif^2 \bm{x}
 \left\{ (\grad^i \eta)^2 + \bar Q_s^2 \eta^2{\rme}^{\eta}
 \right\}\,,\eeq
which involves an {\em exponential} potential. Clearly, this is
now an interacting field theory in which the parameter $\sigma$
--- which, we recall, is a measure of the dispersion introduced by
fluctuations --- plays the role of a coupling constant, as it can
be recognized after rescaling $\eta\equiv\sigma\phi$ :
 \beq\label{eq-S1phi}
 S_1[\phi]= \int \dif^2 \bm{x}
 \left\{ \frac{1}{2}(\grad^i \phi)^2 +
 \frac{1}{2}\,\bar Q_s^2 \phi^2{\rme}^{\sigma\phi}
 \right\}\,.\eeq
Strong interactions for the field $\phi$ correspond to strong
correlations between spots with different sizes and at different
locations, as physically expected at sufficiently high energy.
However, it should be clear that the {\em precise} form of the
potential for $\phi$, beyond the exponential factor, cannot be
uniquely fixed by the uncertainty principle alone: a factor of
$\phi$ counts like $\ln Q_s$ and hence it cannot modify the
power--counting argument that $\grad^2\sim Q_s^2$. Thus, the
uncertainty principle alone would allow for any potential where
${\rme}^{\sigma\phi}$ is multiplied by an arbitrary polynomial in
$\phi$.

At this point we shall invoke {\em conformal symmetry} to further
specify the form of the potential. As mentioned in the
Introduction, the high--energy evolution equations in QCD in the
leading logarithmic approximation are invariant under scale, and,
more generally, (special) conformal transformations. It is likely
that a similar symmetry should also hold for the correlations
generated by this evolution, at least within limited ranges. This
is a very strong constraint on the effective theory, which almost
uniquely fixes its structure, as we shall explain in what follows.

\subsection{Liouville field theory in a nutshell}
Let us temporarily assume that the effective theory should have
{\em exact} conformal symmetry (this assumption cannot right, as
we shall later argue, but it allows us to provide a first
iteration for the effective action). Then, the potential for
$\phi$ is necessarily a pure exponential, and the corresponding
effective action is the same as the {\em classical Liouville
action\footnote{The classical Liouville theory was extensively
studied at the end of the nineteenth century in connection with
the uniformization problem for Riemann surfaces (see, e.g., the
discussion in \cite{GinMoore}).}} \cite{Se90,Tesch} :
 \beq\label{eq-SL}
 S_L[\phi]= \int \dif^2 \bm{x}
 \left\{ \frac{1}{2}(\grad^i \phi)^2 +
 \frac{Q_0^2}{\sigma^2}\,{\rme}^{\sigma\phi}
 \right\}\,.\eeq
A factor $1/\sigma^2$ has been introduced in the coefficient of
the potential to ensure that, in the weak--coupling limit
$\sigma\to 0$ with fixed $\eta\equiv\sigma\phi$, both terms in the
action are of $\mcal{O}(1/\sigma^2)$. Also, for later convenience,
we have replaced the mass scale $\bar Q_s^2$ in front of the
potential by the generic scale $Q_0^2$. Correspondingly, we
re-interpret the field $\phi$ as
 \beq\label{eq-phidef}
 \phi(\bm{x})\,\equiv\,\frac{1}{\sigma}\,\ln\frac {Q_s^2(\bm{x})}
 {Q_0^2}\,,\eeq
so that the quantity
${Q_0^2}\,{\rme}^{\sigma\phi(\bm{x})}=Q_s^2(\bm{x})$ preserves its
original meaning as the event--by--event saturation momentum (a
composite, `vertex', operator in the present field theory).

The exponential potential has the remarkable property to be
invariant under conformal transformations provided one allows the
field $\phi$ to change by a shift under such transformations.
Consider indeed the scale transformation $\bm{x} \to \lambda
\bm{x}$ with $\lambda > 0$. Then the action (\ref{eq-SL}) is
invariant under the following transformations:
 \beq\label{eq-lambda}
 \bm{x} &\,\to\,& \bm{x}' = \lambda \bm{x}\nn
 \phi(\bm{x})&\,\to\,& \phi'(\bm{x}')\,=\,\phi(\bm{x})\,-\,
 \frac{2}{\sigma}\,\ln\lambda\,.
 \eeq
More generally, $S_L$ is invariant under {\em local} conformal
transformations, that is, the general holomorphic transformations
of the complex plane (which include the special, or global,
conformal transformations). To formulate this symmetry, it is
convenient to introduce complex notations: $\bm{x}=(x_1,x_2) \to z
= x_1 + \rmi x_2$. Then, the action (\ref{eq-SL}) is invariant
under
  \beq\label{eq-conf}
 z &\,\to\,& w=f(z)\nn
 \phi(z)&\,\to\,& \phi'(w)\,=\,\phi(z)\,-\,
 \frac{1}{\sigma}\,\ln\left|\frac{\del f}{\del z}\right|^2\,.
 \eeq
since, e.g., $\rme^{\sigma\phi}\dif z\dif \bar
z=\rme^{\sigma\phi'}\dif w\dif \bar w$.

Via an appropriate quantization procedure, which turns out to be
quite non--trivial \cite{Se90,Tesch}, the above symmetry
properties can be carried over to the {\em quantum} version of the
Liouville field theory (QLFT), with important consequences: First,
being invariant under a continuum class of symmetry
transformations, the quantum Liouville theory is {\em integrable}.
Second, the point--dependence of the correlation functions of the
vertex operator $V(\bm{x})\equiv{\exp}\{\sigma\phi(\bm{x})\}$ is,
to a large extent, fixed by the Ward identities associated with
conformal transformations (see, e.g., the textbook discussion in
\cite{CFT}). This yields, e.g.,
   \beq\label{eq-VL}
  \langle V(\bm{x})V(\bm{y})\rangle&\,=\,&
  \frac{C_2(\sigma)}
  {|\bm{x-y}|^{4}Q_0^{4}}
  \,,\nn
  \langle V(\bm{x})V(\bm{y})V(\bm{z})\rangle&\,=\,&
    \frac{C_3(\sigma)}
  {|\bm{x-y}|^{2}|\bm{y-z}|^{2}|\bm{z-x}^{2}|Q_0^6}
\,,
 \eeq
where the coefficients $C_2(\sigma)$, $C_3(\sigma)$, etc., are
known {\em exactly} \cite{DO94,ZZ96}. These formul\ae{} follow
from the conformal symmetry of the (properly defined) path
integral together with the fact that $V(\bm{x})$ is a primary
field with scaling dimension $\Delta=1$ (cf.
Eq.~(\ref{eq-lambda})) :
 \beq\label{eq-Vlambda}
 \bm{x} \,\to\,\bm{x}' = \lambda \bm{x}\quad\Longrightarrow\quad
 V(\bm{x})&\,\to\,& V'(\bm{x}')\,=\, \frac{1}{\lambda^2}\,V(\bm{x})\,
 \equiv\,\frac{1}{\lambda^{2\Delta}}\,V(\bm{x})\,.
 \eeq
This particular transformation law for the vertex operator is
natural from the point of view of  QCD: under a scale
transformation, the saturation momentum $Q_s^2(\bm{x})\equiv
{Q_0^2}V(\bm{x})$ transforms as an operator with mass dimension
two, which is its physical dimension indeed.

A rather subtle aspect of LFT, which complicates its quantum
implementation, refers to the interplay between ultraviolet
renormalization and conformal symmetry: although
superrenormalizable (since the exponential potential can be
expanded out in a series in powers of $\phi$ and each term in this
series is superrenormalizable in $d=2$), the quantum theory
requires renormalization to remove `tadpole' divergences
--- i.e., divergences associated with the equal--point limit of
the propagator. Such divergences arise from contracting fields at
the same point and thus can be eliminated by normal--ordering the
operators. But this procedure introduces `anomalous dimensions'
for the vertex operators (their scaling dimensions acquire quantum
corrections), which could spoil conformal symmetry. To maintain
the symmetry, the theory is modified in such a way to ensure that
the vertex operator $V=\rme^{\sigma\phi}$ which enters the action
preserves at quantum level the classical dimension $\Delta=1$, cf.
Eq.~(\ref{eq-Vlambda}). Note however that for a generic vertex
operator $V_\alpha\equiv\rme^{\alpha\phi}$ with $\alpha\ne\sigma$,
the ensuing quantum dimension is different from the respective
prediction $\Delta_\alpha^{\rm (cl)}=\alpha/\sigma$ of the
classical Liouville theory (cf. Eq.~(\ref{eq-lambda})) (see, e.g.,
\cite{Se90,ZZ96,Tesch}). Here, we do not need to discuss these
complications in more detail because, as we shall see, they do not
show up in the modified version of the Liouville theory that we
shall propose as an effective theory for the QCD problem at hand.
In fact, the only reason for us to mention these subtleties here,
it is to emphasize, by contrast, the situation in our final theory
(cf. Sect. \ref{SECT-EFT}), where all such UV complications
disappear.

But before turning to that presentation, let us rapidly explain
why the standard QLFT cannot be, by itself, the complete effective
theory that we need in QCD. The problem comes precisely from that
feature of this theory which is also its main virtue : the exact
conformal symmetry. Because of this symmetry, QLFT involves no
real mass scale --- the scale $Q_0^2$ apparent in
Eq.~(\ref{eq-SL}) has no intrinsic meaning since its magnitude can
be changed at will (and thus made arbitrarily large or arbitrarily
small) by shifting the field $\phi$ under the path integral ---,
and hence it cannot describe a non--trivial gluon distribution, as
characterized by a non--zero expectation value for the saturation
momentum. In fact, the Ward identities for conformal symmetry
\cite{CFT} imply $\langle V(\bm{x})\rangle=0$, showing that, from
the perspective of QCD, the LFT would be a theory for
fluctuations, but ... without matter: $\langle
Q_s^2(\bm{x})\rangle=0$ at any $\bm{x}$ in pure QLFT !

The same basic problem can be seen under different angles,
revealing as many `paradoxes' from the point of view of QCD: The
exponential potential in Eq.~(\ref{eq-SL}) has no local minimum
and becomes flat when $\phi$ is negative and large; hence, the
classical field is rolling down to $-\infty$ and, correspondingly,
the quantum theory has no stable ground state. Because of that, it
makes no sense to compute the correlation functions of the field
$\phi$ --- only the correlations of the vertex operator
$V_\alpha\equiv\rme^{\alpha\phi}$ (with generic $\alpha$), or the
mixed correlations involving both $\phi$ and $V_\alpha$, are {\em
a priori} well defined. Still because of the lack of a stable
ground state, there is no fundamental difference between `weak'
and `strong' coupling in QLFT:  the vertex correlation functions
exhibit the same power--law behaviour, cf. Eq.~(\ref{eq-VL}), for
{\em any} value of $\sigma$, small or large, since this behaviour
is fixed by conformal symmetry alone. Besides, the coefficients in
these correlations, so like $C_2(\sigma)$ and $C_3(\sigma)$ in
Eq.~(\ref{eq-VL}), show interesting `self--duality' properties
under the exchange $\sigma\leftrightarrow 1/\sigma$ \cite{ZZ96}.
There is furthermore no distinction between `short' and `long'
distances, precisely because there is no intrinsic mass scale in
the theory: QLFT generates power--law correlations for the vertex
operators on {\em all} scales.

Clearly, all these properties would be unacceptable in the
framework of our original QCD problem, where, on the contrary, we
expect pronounced differences between the weak and the strong
coupling regimes, or between short and large distances. To
understand the way out of such paradoxes, let us remind that, in
the context of QCD, the conformal symmetry characteristic of the
evolution equations is explicitly broken by the initial condition
at $Y=0$, which introduces a physical scale in the problem. It is
the subsequent evolution of this scale with increasing energy
which fixes the average saturation momentum at some later $Y$. But
the effective theory that we are looking for is not a theory for
the evolution, but rather for the {\em results} in this evolution
(in terms of correlations of $Q_s$) at the final rapidity $Y$. In
this theory, the average saturation momentum at $Y$ is a parameter
that must be introduced by hand, with the effect that conformal
symmetry is explicitly broken. A particularly simple
implementation of this idea will be described in the next section.

\subsection{Liouville field theory with a homogeneous source term}
\label{SECT-EFT}

From now on, we shall adopt the point of view that the Liouville
action describes the {\em fluctuations} in the high--energy
evolution (since it has the correct symmetry in that sense), but
that in order to also describe the {\em average} gluon
distribution, this action must be supplemented with a `source
term' which breaks down the scale symmetry and thus provides a
non--zero expectation value for the saturation momentum. We shall
consider here the simplest source term, namely, an operator linear
in $\phi$ whose strength (the source density) is adjusted in such
a way to produce a prescribed value for the average saturation
momentum. For simplicity, we shall first consider a {\em
homogeneous} gluon distribution, as produced by the high--energy
evolution of an initial distribution which was itself homogeneous
(say, a `large nucleus' in the context of the
McLerran--Venugopalan model \cite{MV}). This is not a very strong
restriction, since the typical length scales that we shall be
interested in at high energy are anyway much shorter than the
scales characterizing the inhomogeneity in a physical hadron at
low energy.

We thus introduce the following extension of the Liouville action:
$S=S_L + j\int_{\bm{x}}\phi(\bm{x})$, where $S_L$ is given by
Eq.~(\ref{eq-SL}) and $j$ is the source density, assumed to be
homogeneous. There are several ways to fix $j$, all of them
leading to the same result. For instance, by requiring $S$ to
reduce to the Gaussian action (\ref{eq-S0}) in the limit of small
fluctuations/weak coupling $\sigma\phi\ll 1$, one immediately
finds that the source term should remove the linear term from the
expansion of the Liouville exponential. One thus finds:
  \beq\label{eq-OURS}
 S[\phi]= \int \dif^2 \bm{x}
 \left\{ \frac{1}{2}(\grad^i \phi)^2 +
 \frac{Q_0^2}{\sigma^2}\,\Big({\rme}^{\sigma\phi}  - \sigma\phi
 -1\Big)
 \right\}\,,\eeq
where we have also subtracted the constant, unit, term from the
exponential, for convenience. By construction, the above action
has the saddle point $\phi=0$, which controls the dynamics in the
weak coupling (i.e., low energy) regime $\sigma\ll 1$. More
generally, the above potential has a unique minimum at $\phi=0$,
which implies that the quantum theory defined by $S$ has a stable
ground state (at least, in perturbation theory). In particular,
within this theory it makes sense to perturbatively compute the
correlation functions of $\phi$ (unlike in LFT).

The comparison between Eq.~(\ref{eq-OURS}) and the quadratic
action in Eq.~(\ref{eq-S0}) seems to suggest $Q_0^2 \equiv\bar
Q_s^2$. However, this identification is wrong in general --- it
holds only as an approximate equality in the weak--coupling regime
where the action (\ref{eq-OURS}) is supposed to reduce to
Eq.~(\ref{eq-S0}). To make the proper identification in the
general case, consider the first Dyson equation generated by
Eq.~(\ref{eq-OURS}), that is
    \beq\label{eq-Dyson}
  \lan \frac{\delta S}{\delta
  \phi(\bm{x})}\ran=\,0\quad\Longleftrightarrow\quad
  -\grad^2\!\lan\phi(\bm{x})\ran +
  \frac{Q_0^2}{\sigma}\,\lan
  {\rme}^{\sigma\phi}\ran=\,\frac{Q_0^2}{\sigma}\,.
  \eeq
By homogeneity, the mean field $\lan\phi(\bm{x})\ran$ is
independent of $\bm{x}$, hence $\grad^2\!\lan\phi\ran =0$, and
then Eq.~(\ref{eq-Dyson}) implies:
    \beq\label{eq-V1}
 \lan {\rme}^{\sigma\phi}\ran\,=\,1\,,\qquad\mbox{or}\qquad
  \langle Q_s^2 \rangle\,\equiv\,Q_0^2 \,\lan
 {\rm e}^{\sigma\phi}\ran\,=\,Q_0^2\,.
 \eeq
Thus, the scale $Q_0^2$ in the action (\ref{eq-OURS}) must be
understood as the {\em average saturation momentum} in the sense
of Eq.~(\ref{eq-QSdef2}) (and not of Eq.~(\ref{eq-QSdef1}) !). In
fact, one could have alternatively determined the value of the
source density $j$ by directly requiring $\langle Q_s^2 \rangle
=Q_0^2$ (rather than going through a weak coupling argument); via
the first Dyson equation, this condition would have again implied
$j=-Q_0^2/\sigma$, as above.

Note the importance of the source term in the action for the
previous argument: without that term, the r.h.s. of
Eq.~(\ref{eq-Dyson}) would be zero, and then the only solution
consistent with homogeneity would be $\langle
{\rme}^{\sigma\phi}\rangle=0$, as expected in QLFT. The source
term explicitly breaks down conformal symmetry and thus introduces
a mass scale in the problem, to be physically identified with the
average value of the saturation momentum. To better appreciate the
role of the source term in that sense, imagine starting with two
different scales, say, $M^2$ (instead of $Q_0^2$) in front of the
Liouville potential in Eq.~(\ref{eq-SL}) and $Q_0^2$ in front of
the source term. After shifting the field as
$\phi=\tilde\phi+\phi_0$, with $\sigma\phi_0\equiv
\ln(Q_0^2/M^2)$, the scale $M^2$ gets replaced by $Q_0^2$ in the
action for $\tilde\phi$. This `scale transmutation' is generic in
relation with LFT: since the Liouville potential is by itself
scale--invariant, the associated mass parameter ($M^2$ in the
above discussion) adjusts itself to the scale introduced by the
symmetry--violating term, if any.

Eq.~(\ref{eq-V1}) is remarkable also in a different respect: it
shows that, in the present theory, the one--point function
$\langle {\rme}^{\sigma\phi}\rangle$ is {\em finite} (and equal to
one) without any ultraviolet renormalization. This is remarkable
since, {\em a priori}, one would expect this quantity to be
dominated by the fluctuations with the highest momenta and thus be
afflicted with ultraviolet divergences. This was already the case
in the free field theory, cf. Eq.~(\ref{eq-Vren}), and this is
also the general situation in the known two--dimensional field
theories, which are superrenormalizable, but not strictly finite.
The finite result in Eq.~(\ref{eq-V1}) anticipates a more general
property of the effective theory (\ref{eq-OURS}), that we shall
verify in Sect. \ref{SECT_UV} via explicit perturbative
calculations up to two--loop order\,: Namely, this theory is {\em
ultraviolet finite}, in the following sense: all the $n$--point
functions (with $n\ge 2$) of the Liouville field $\phi$, as well
as all the $n$--point correlation functions (with $n\ge 1$) of the
vertex operator $V=\rme^{\sigma\phi}$, come out truly finite when
computed in perturbation theory and thus do not require
ultraviolet renormalization. (More general operators, however,
like $\rme^{\alpha\phi}$ with $\alpha\ne\sigma$, can still meet
with UV divergences, which then can be renormalized in the
standard way.) As we shall discover in Sect. \ref{SECT_UV}, this
UV--finiteness comes out as a result of order--by--order
cancellations between divergent, `tadpole', diagrams, which are
abundantly produced by the perturbative expansion, but which
precisely cancel with each other, due to the special symmetry
factors of the interaction vertices. An important consequence of
such cancellations is that the vertex operator has no anomalous
dimension.

But whereas the perturbation theory for the effective theory is
meaningful and rather straightforward (see Sect. \ref{SECT_UV}),
it is on the other hand difficult to derive firm results about the
non--perturbative behaviour of the theory in the interesting
regime at high energy, or strong coupling, $\sigma\gg 1$ (with the
noticeable exception of Eq.~(\ref{eq-V1})). Unlike in the standard
Liouville theory, here one cannot rely anymore on conformal
symmetry to deduce, or at least constrain, the general form of the
correlations. In what follows we shall attempt to deduce some
general properties of the theory from a qualitative analysis of
its action (\ref{eq-OURS}).

Since this action involves a physical mass scale $Q_0^2$, which is
moreover the curvature of the potential at its minimum, it is
quite clear that the correlation functions in this theory should
die out exponentially over sufficiently large distances $R\gg
1/Q_0$. However, the potential in Eq.~(\ref{eq-OURS}) is not a
standard mass term for $\phi$, and this difference has interesting
consequences:

First, the minimum of the potential, which roughly speaking
indicates the most probable value for the saturation momentum,
occurs at a harder scale in Eq.~(\ref{eq-OURS}) than it was the
case in the free field theory (\ref{eq-S0}), or in the Gaussian
approximation (\ref{eq-prob}). Indeed, in Eq.~(\ref{eq-OURS}) this
minimum corresponds to $Q_s^2=Q_0^2 \equiv Q_0^2 \langle {\rm
e}^{\sigma\phi}\rangle$, whereas in Eqs.~(\ref{eq-S0}) and
(\ref{eq-prob}) it rather corresponds to $Q_s^2=\bar Q_s^2\equiv
{Q_0^2}\,{\rme}^{\sigma\langle\phi\rangle}$ (recall
Eqs.~(\ref{eq-QSdef1})--(\ref{eq-QSdef2})). Since $\langle {\rm
e}^{\sigma\phi}\rangle$ is always larger than
${\rme}^{\sigma\langle\phi\rangle}$, it is clear that, in the
theory with action (\ref{eq-OURS}), the fluctuations in $Q_s^2$
are pushed towards harder scales, as anticipated.

Second, unlike the quadratic potential in Eqs.~(\ref{eq-S0}) or
(\ref{eq-prob}), the one in Eq.~(\ref{eq-OURS}) is asymmetric
under $\phi\to -\phi$, and this asymmetry is very pronounced for
the relatively strong fluctuations with $\sigma\phi\simge 1$: the
potential favors large {\em negative} fluctuations as opposed to
large {\em positive} ones. One may think that the exponential
piece of this potential totally forbids the fluctuations with
$\sigma\phi \simge 1$ (or $Q_s^2\gg Q_0^2$), but this is actually
not so: arbitrarily hard fluctuations with $Q_s^2\gg Q_0^2$ are
still allowed, because they have tiny sizes and thus give small
contributions to the action. Such a propensity towards small--size
fluctuations is of course natural in any field theory, but this is
not properly taken into account by the coarse--graining
approximation (\ref{eq-prob}), which ignores any information about
the sizes of the spots.

To be (slightly) more quantitative, let us estimate the
contribution of a fluctuation with size $R$ to the action. The
typical gradients for this configuration are $\grad \sim 1/R$,
hence the kinetic term contributes $\int_{R} (\grad^i \phi)^2 \sim
R^2 (\grad^i \phi)^2 \sim \phi^2$. After similarly estimating the
potential term, one has
 \beq\label{eq-SR}
 S(R) \,\sim\, \phi^2 \,+ R^2\,\frac{Q_0^2}{\sigma^2}\,
 \big({\rme}^{\sigma\phi}\,  -
 \,{\sigma}{\phi} \big)
 \,.\eeq
Roughly speaking, the allowed fluctuations are those for which
$S(R)\simle 1$.

Consider first the situation in pure Liouville theory, i.e.,
without the source term; then, whatever the value of $R$ is, the
exponential potential allows for all the fluctuations with $\phi
\simle \phi_{\rm max}$, with $\sigma\phi_{\rm max} \equiv \ln(
\sigma^2/R^2Q_0^2)$. Note that $\sigma\phi_{\rm max}$ can be
arbitrarily large, as anticipated, provided $R$ is correspondingly
small. The correlation functions of the vertex operator are
dominated by the fluctuations with maximal field strength, which
implies, e.g.,
 \beq\label{eq-VL2}
  \langle {\rme}^{\sigma\phi(\bm{x})}{\rme}^{\sigma\phi(\bm{y})}
  \rangle\,\sim\,
  \frac{\sigma^4}
  {R^{4}Q_0^{4}}\qquad\mbox{where}\qquad |\bm{x-y}| = R,\eeq
in qualitative agreement (in so far as the $R$--dependence is
concerned) with the correct result, Eq.~(\ref{eq-VL}). Note also
that the saturation momentum for the relevant fluctuations is
correlated to their size, ${Q_0^2}{\rme}^{\sigma\phi_{\rm max}}
\simge 1/R^2$, in agreement with the uncertainty principle.

We now turn to the full potential, including the linear source
term. Then, we need to distinguish between two kinds of
fluctuations --- small--size and large--size ---, according to the
value of the ratio $R^2Q_0^2/\sigma^2$ :

\texttt{(i)} When $R^2Q_0^2\ll \sigma^2$, we expect the same
situation as in pure Liouville theory\footnote{We implicitly
assume here that the coupling is sufficiently strong:
$\sigma\simge 1$.}. Indeed, in this case, the potential in
Eq.~(\ref{eq-SR}) authorizes fluctuations within the relatively
wide range $\phi_{\rm min} \simle \phi \simle \phi_{\rm max}$,
with $\phi_{\rm max}$ as above and ${\sigma}\phi_{\rm min}\equiv -
\sigma^2/R^2Q_0^2$. The correlations of ${\rme}^{\sigma\phi}$ are
controlled by the fields towards the upper limit, where the source
term is negligible; that is, they are determined by the Liouville
piece of the action, and thus are power--like, with the same
powers as in LFT.

\texttt{(ii)} On the other hand, for $R^2Q_0^2\simge \sigma^2$,
the potential allows only weak--amplitude fluctuations, such that
${\sigma}|\phi|\simle\sigma^2/R^2Q_0^2\simle 1$. (By itself, the
exponential piece of the potential would also allow for larger
negative values, but these are suppressed by the source term.)
Within this range, the potential reduces to a quadratic mass term
with mass $Q_0^2$. Hence, the large--size fluctuations die out
exponentially over a typical distance $1/Q_0$.

To summarize, the (admittedly crude) estimates above suggest that,
in the effective theory with action (\ref{eq-OURS}), the
correlations of the vertex operator ${\rm e}^{\sigma\phi}$ have a
power--law behaviour, with Liouville--like exponents, over short
distances $R\simle 1/Q_0$, but they decay exponentially over
larger distances $R\gg 1/Q_0$. Note that, in order to conclude in
favor of Liouville--like exponents at short distances, it was
essential that the vertex operator has no anomalous dimension, as
we shall check via perturbative calculations in Sect.
\ref{SECT_UV}.

\subsection{A more general source term}
\label{SECT-SOURCE}

A physical hadron is never homogeneous, and the original
inhomogeneity at low energy gets transmitted to, and it is
modified by, the high energy evolution. Assume, e.g., that one
starts with a large nucleus at $Y=0$, in which the gluon density
is large and quasi--homogeneous inside a large disk of radius $R$,
but it rapidly drops out to zero at impact parameters larger than
$R$ (so like in the McLerran--Venugopalan model \cite{MV}).
Whatever was the initial law for this fall--off at large
distances, it will get replaced by a {\em power} law after
evolving the system to sufficiently high energies according to the
perturbative evolution equations in QCD. The BK equation (or,
equivalently, the BFKL equation supplemented with a saturation
boundary condition) predicts that, for the saturation momentum,
this power must be 4: $Q_s^2(\bm{x})\sim 1/x^4$ for $x\gg R$ (see,
e.g., \cite{MW03,GBS03}). It is therefore interesting to consider
generalizations of the effective theory introduced in the previous
subsection which allow for a inhomogeneous source term. The action
then becomes $S=S_L + \int_{\bm{x}} j(\bm{x})\phi(\bm{x})$, with
$j(\bm{x})$ describing the strength and the impact parameter
dependence of the average saturation momentum.

The functional form of $j(\bm{x})$ is in principle fixed by the
underlying evolution equations (together with the initial
conditions at low energy) and represents a `free parameter' from
the perspective of the effective theory. Here, we shall
parameterize this in the form
  \beq\label{eq-jphi0}
 j(\bm{x})\,=\,- \frac{Q_0^2}{\sigma}\,
 {\rme}^{\sigma\phi_0(\bm{x})}\,,\eeq
where $Q_0^2(Y)$ determines the value of the average saturation
momentum at the center of the hadron and at rapidity $Y$, whereas
the function $\phi_0(\bm{x})$, with $\phi_0(0)=0$, describes the
profile of $\langle Q_s^2(\bm{x})\rangle$ in impact--parameter
space. As previously mentioned, a physically motivated choice is
  \beq\label{eq-phi0x}
 {\rme}^{\sigma\phi_0(\bm{x})}\,=\,\frac{1}{[1+x^2/R^2]^2}
 \,,\eeq
with $x=|\bm{x}|$ and $R$ the hadron radius at $Y=0$. The current
in Eq.~(\ref{eq-jphi0})--(\ref{eq-phi0x}) is clearly only an
approximation, as it implicitly assumes that the evolutions in $Y$
and in $\bm{x}$ decouple from each other; yet, this approximation
captures the salient features of this evolution, namely the fact
that the (average) saturation momentum rises rapidly with $Y$ at
any $x$ and it develops a $1/x^4$--tail at distances $x\gg R$. We
shall furthermore assume that $Q_0^2(Y) R^2 \gg 1$ at any $Y$, as
appropriate for a sufficiently large nucleus which is relatively
dense already at $Y=0$ and has a size $R$ fixed by the
non--perturbative, soft, physics.

The first Dyson equation corresponding to this current, that is
(cf. Eq.~(\ref{eq-Dyson}))
    \beq\label{eq-Dysonj}
  -\grad^2\!\lan\phi(\bm{x})\ran +
  \frac{Q_0^2}{\sigma}\,\lan
  {\rme}^{\sigma\phi(\bm{x})}\ran=\,\frac{Q_0^2}{\sigma}\,
 {\rme}^{\sigma\phi_0(\bm{x})}\,,
  \eeq
cannot be exactly solved in general, but it clearly implies
$\langle Q_s^2(\bm{x})\rangle\approx Q_0^2\,
{\rme}^{\sigma\phi_0(\bm{x})}$. (Indeed, the scale of the
inhomogeneity being fixed by $R$, one has $|\grad^2\!\lan\phi\ran|
\sim R^2$, which is much smaller than ${Q_0^2}/{\sigma}$ at any
$Y$.)

It is furthermore quite clear that the short--distance behaviour
of the theory (on distance scales $r\ll R$) cannot be changed by
the soft inhomogeneity visible in Eq.~(\ref{eq-phi0x}). In
particular, the relevant correlation functions (those of the
Liouville field and of the vertex operator
${\rme}^{\sigma\phi(\bm{x})}$) are still ultraviolet finite. Also,
these correlations preserve the same behaviour on short ($r\simle
1/Q_0$) and intermediate ($1/Q_0 \ll r\ll R$) distances as in the
homogeneous case. On the other hand, on very large distances $r\gg
R$, the correlation should show a slower decay, although still
exponential, because the effective mass for this decay, namely
$\langle Q_s^2(\bm{x})\rangle$, becomes smaller and smaller when
increasing the distance from the origin.

The role of $\langle Q_s^2(\bm{x})\rangle$ as an effective mass
becomes manifest on the expansion of the action $S$ around its
saddle point at $\sigma\to 0$, as appropriate for the purposes of
perturbation theory. Specifically, the saddle point condition is
obtained by removing the brackets in Eq.~(\ref{eq-Dysonj}) and
determines the classical solution $\phi_{\rm cl}(\bm{x})$ :
 \beq\label{eq-saddle}
  -\grad^2\,\phi_{\rm cl}(\bm{x}) +
  \frac{Q_0^2}{\sigma}\,
  {\rme}^{\sigma\phi_{\rm cl}(\bm{x})}=\,\frac{Q_0^2}{\sigma}\,
 {\rme}^{\sigma\phi_0(\bm{x})}\,.
  \eeq
For instance, for the particular current (\ref{eq-phi0x}), the
above equation can be easily solved to give
 \beq\label{eq-phicl}
 {\rme}^{\sigma\phi_{\rm cl}(\bm{x})}\,=\,\frac{C}{[1+x^2/R^2]^2}
 \qquad \mbox{with}\qquad C = 1 - \frac{8}{Q_0^2 R^2}\,\approx 1
 \,.
 \eeq
By separating $\phi=\phi_{\rm cl}+\delta\phi$ and expanding around
$\phi_{\rm cl}$, one finds the action which governs the dynamics
of the fluctuation field $\delta\phi$:
  \beq\label{eq-deltaphi}
 S[\phi]&\,=\,& S[\phi_{\rm cl}]+\int \dif^2 \bm{x}
 \left\{ \frac{1}{2}(\grad^i \delta\phi)^2 +
 \frac{Q_{\rm cl}^2(\bm{x})}
 {\sigma^2}\,\Big({\rme}^{\sigma\delta\phi}  - \sigma\delta\phi
 -1\Big)
 \right\}\nn&\,=\,&
 S[\phi_{\rm cl}]+\int \dif^2 \bm{x}
 \left\{ \frac{1}{2}(\grad^i \delta\phi)^2 +
 \frac{1}{2}{Q_{\rm cl}^2(\bm{x})}\delta\phi^2
 + Q_{\rm cl}^2(\bm{x})\sum_{k\ge 3}\,
 \frac{\sigma^{k-2}\delta\phi^k}{k!}\right\}
 \,,\eeq
where $Q_{\rm cl}^2(\bm{x})\equiv Q_0^2{\rme}^{\sigma\phi_{\rm
cl}(\bm{x})}$ is the value of $\langle Q_s^2(\bm{x})\rangle$ in
the saddle point approximation, and plays the role of a
point--dependent mass for $\delta\phi$, as anticipated. The
expansion in the second line of Eq.~(\ref{eq-deltaphi}) will be
used in the perturbative calculations to be presented in the next
section (for the homogeneous case $\phi_0=0$, for simplicity).

\section{Ultraviolet finiteness of the correlation functions}
\setcounter{equation}{0}\label{SECT_UV}

In this section, we shall demonstrate via explicit calculations
that some interesting classes of correlations generated by the
effective theory are ultraviolet finite in perturbation theory.
Our calculations will be performed only to finite orders (namely,
up to two--loop order), and thus they cannot be seen as a
complete, and even less rigorous, proof in that sense. Yet, as we
shall see, they reveal a very non--trivial pattern of tadpole
cancellations, which is very unlikely to be accidental, but most
probably is representative for the way how such cancellations
proceed to all orders in the perturbative expansion.

The correlations that we shall show to be finite are the
$n$--point {\em connected} correlation functions of $\phi$ with
$n\ge 2$ and all the $n$--point correlation functions (with $n\ge
1$) of the vertex operator $V=\rme^{\sigma\phi}$. The mean field
$\langle\phi\rangle$, on the other hand, appears {\em not} to be
finite, as its perturbative expansion starts with a divergent
contribution of $\mcal{O}(\sigma)$ (while the respective
corrections of higher orders are still found be finite). But this
unique divergence plays an essential role in that it cancels, via
its iterations, similar divergences which appear in the expansion
of the vertex operator and in the disconnected pieces of the
$n$--point functions of $\phi$.

\subsection{The $n$--point functions of the Liouville field}
\label{SECT_CorrelPhi}

We shall first consider the $n$--point functions of the Liouville
field $\phi$, which are interesting not only by themselves, but
also as ingredients in the respective calculations for the vertex
operator, to be presented in the next subsection.

As a representative example, we shall consider the perturbative
expansion for 2--point function $\langle
\phi(\bm{x})\phi(\bm{y})\rangle$ up to $\mcal{O}(\sigma^4)$,
meaning two--loop order for the self--energy. Our presentation
will focus on the systematics of tadpole cancellations, which is
our main concern here.

The expectation value is defined by the following path integral:
   \beq\label{eq-phi2}
 \lan \phi(\bm{x})\phi(\bm{y})\ran\equiv \,\frac{1}{Z}
 \int D[\phi]\,\,\phi(\bm{x})\phi(\bm{y})\,
 \rme^{-S[\phi]}\,,
 \eeq
where $S$ is the action in Eq.~(\ref{eq-OURS}) and $Z$ denotes the
partition function: $Z\equiv \int
D[\phi]\,\exp\left\{-S[\phi]\right\}$. For the purposes of the
perturbation theory, we separate the quadratic and the interaction
parts of $S$ in the standard way: $S=S_0+S_{\rm int}$, where
   \beq\label{eq-Sf}
  S_0[\phi]\equiv \,\frac{1}{2}\int \dif^2 \bm{x}
 \left\{ (\grad^i \phi)^2 + Q_0^2 \phi^2\right\}\,,\eeq
provides the free propagator $D$ (that is, Eq.~(\ref{eq-D0}) with
$\bar Q_s^2\to Q_0^2$), and
 \beq\label{eq-Sint}
  S_{\rm int}[\phi]\equiv \,\frac{Q_0^2}{\sigma^2}\int \dif^2 \bm{x}
 \,\left\{{\rme}^{\sigma\phi} -1 - \sigma\phi
 -\frac{1}{2}(\sigma\phi)^2
 \right\} =\,\sigma Q_0^2\int \dif^2 \bm{x} \sum_{k\ge 3}\,
 \frac{\sigma^{k-3}\phi^k}{k!}
 \,,\eeq
provides the interaction vertices. The special symmetry factors
associated with these vertices, as generated by the expansion of
the exponential, lie at the heart of the ultraviolet finiteness to
be demonstrated below: at any given order in perturbation theory,
tadpoles produced by various vertices $\phi^k$ cancel with each
other, because of these special symmetry factors.

The following identity, involving the free propagator, will be
also useful in what follows :
  \beq\label{eq-intD0}
 \int_{\bm{z}}\,D(\bm{x}-\bm{z})
 \equiv \int \dif^2 \bm{z}\int\frac{\dif^2 \bm{k}}{(2\pi)^2}
 \,\frac{\rme^{i\bm{k}\cdot \bm{(x-z)}}}{k^2+Q_0^2}\,
 \,=\,\frac{1}{Q_0^2}\,.\eeq

\begin{figure}
\begin{center}
\centerline{\epsfig{file=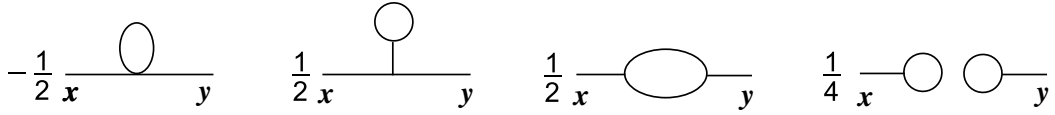,height=1.6cm}}
 \caption{The four diagrams contributing to
the 2--point function $\lan \phi(\bm{x})\phi(\bm{y})\ran$ to
$\mcal{O}(\sigma^2)$, in the order in which they are listed in
Eq.~(\ref{eq-phi23}).} \label{P2P1L}
\end{center}
\end{figure}

By expanding $S_{\rm int}$ in Eq.~(\ref{eq-phi2}) to
$\mcal{O}(\sigma^2)$, one finds $\lan \phi(\bm{x})\phi(\bm{y})\ran
\approx D(\bm{x},\bm{y})+\lan \phi(\bm{x})\phi(\bm{y})\ran^{(2)}$,
with (an upper index on an expectation value indicates the order
in $\sigma$ to which the respective expectation value is to be
evaluated)
 \beq\label{eq-phi22}
 \lan \phi(\bm{x})\phi(\bm{y})\ran^{(2)}&\,=\,&
 -\frac{\sigma^2Q_0^2}{4!}
 \int_{\bm{z}}\,\lan\phi(\bm{x})\phi(\bm{y})
 \phi^4(\bm{z})\ran_0\nn &\,+\,&
 \frac{\sigma^2Q_0^4}{2(3!)^2}
 \int_{\bm{z}_{1},\bm{z}_2}\,\lan\phi(\bm{x})\phi(\bm{y})
 \phi^3(\bm{z}_1)\phi^3(\bm{z}_2)\ran_0\,
 \,,
 \eeq
where $\lan\cdots\ran_0\equiv \lan\cdots\ran^{(0)}$ is a simpler
notation for the expectation value computed with the quadratic
action $S_0$. After performing the above contractions, one obtains
the following four terms (recall that the vacuum diagrams are
eliminated by the denominator $Z$ in Eq.~(\ref{eq-phi2}))
 \beq\label{eq-phi23}
 \lan \phi(\bm{x})\phi(\bm{y})\ran^{(2)}&\,=\,&
 -\frac{\sigma^2Q_0^2}{2}\,D(0)\int_{\bm{z}}\,
 D(\bm{x}-\bm{z})D(\bm{z}-\bm{y})
 \nn &\,+\,&\frac{\sigma^2Q_0^4}{2}\,D(0)
 \int_{\bm{z}_{1},\bm{z}_2}\,D(\bm{x}-\bm{z}_1)
 D(\bm{z}_1-\bm{y})D(\bm{z}_1-\bm{z}_2)
 \nn &\,+\,&\frac{\sigma^2Q_0^4}{2}
 \int_{\bm{z}_{1},\bm{z}_2}\,D(\bm{x}-\bm{z}_1)
 D^2(\bm{z}_1-\bm{z}_2)
 D(\bm{z}_2-\bm{y})
  \nn &\,+\,&\left[\frac{\sigma Q_0^2}{2}\,D(0)
 \int_{\bm{z}}\,D(\bm{x}-\bm{z})\right]^2
 \,,
 \eeq
corresponding to the four diagrams exhibited in Fig. \ref{P2P1L}.
The first two diagrams involve the divergent tadpole $D(0)$, but
the corresponding symmetry factors are such that these diagrams
precisely cancel with each other. (Note that the integral over
$\bm{z}_2$ in the second term can be performed using
Eq.~(\ref{eq-intD0}).) The third diagram yields a finite
contribution in $d=2$. Finally, the last term in
Eq.~(\ref{eq-phi23}), which involves a tadpole squared, is
recognized as the disconnected piece
$\lan\phi(\bm{x})\ran\lan\phi(\bm{y})\ran$ of the 2--point
function $\lan\phi(\bm{x})\phi(\bm{y})\ran$ (to this order). One
has indeed:
  \beq\label{eq-phi1}
 \lan \phi(\bm{x})\ran^{(1)} &\,=\,&-
  \frac{\sigma Q_0^2}{3!}
 \int_{\bm{z}}\,\lan\phi(\bm{x})
 \phi^3(\bm{z})\ran_0\nn \,
 &\,=\,&-\frac{\sigma Q_0^2}{2}\,D(0)
 \int_{\bm{z}}\,D(\bm{x}-\bm{z})
 \,=\,-\frac{\sigma}{2}\,D(0)\,.
 \eeq
To summarize, the self--energy to one--loop order is given by the
three connected diagrams in Fig. \ref{P2P1L} (after amputating the
external lines); the first two diagrams are divergent but cancel
with each other, while the third one is finite and represents the
net one--loop contribution. Note that the cancellation of the UV
divergences has occurred between tadpoles generated by two
different interaction vertices: $\phi^3$ and $\phi^4$.

\begin{figure}
\begin{center}
\centerline{\epsfig{file=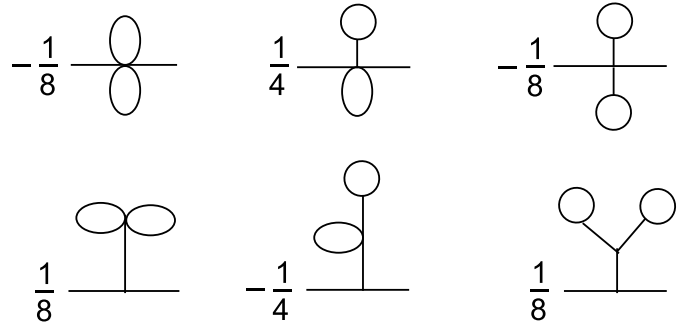,height=4.4cm}} \caption{The
double--tadpole diagrams contributing to the self--energy at
two--loop order.} \label{P2P2T}
\end{center}
\end{figure}

\begin{figure}
\begin{center}
\centerline{\epsfig{file=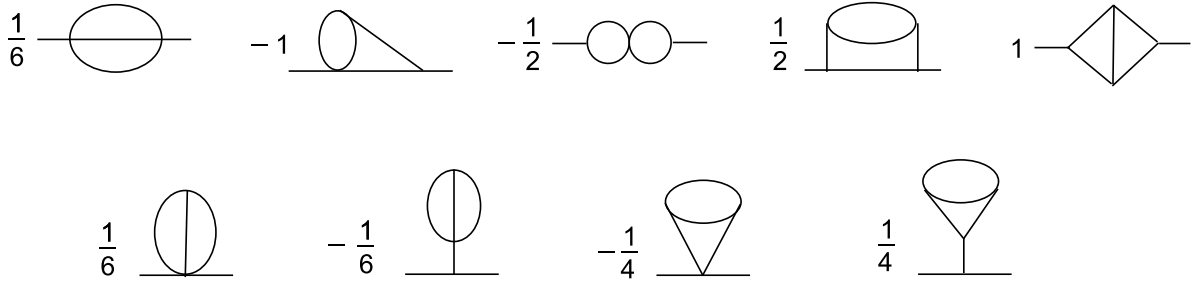,height=3.8cm}} \caption{The
UV--finite two--loop diagrams which give the net result for the
self--energy at $\mcal{O}(\sigma^4)$.} \label{P2P2L}
\end{center}
\end{figure}

When moving to $\mcal{O}(\sigma^4)$, the pattern of tadpole
cancellations becomes significantly more complicated. The
one--loop subdivergences (as associated with self--energy
insertions on the internal propagators) cancel with each other as
explained above, but there are six self--energy diagrams which
feature genuinely two--loop divergences, that is, which are
proportional to $D^2(0)$. These diagrams are displayed in Fig.
\ref{P2P2T}, together with the respective symmetry factors. Each
such a diagram equals $Q_0^2D^2(0)$ times the corresponding
symmetry factor. Thus, as one can read off Fig. \ref{P2P2T}, the
two--loop divergences exactly compensate with each other. The net
contribution to the self--energy to this order is given by the
diagrams exhibited in Fig. \ref{P2P2L}, all of them being finite.
(We do not display this final result, since not particularly
illuminating.) Note that, at this level, the compensating tadpoles
are produced by vertices $\phi^k$ with $k$ ranging from 3 to 6.
Clearly, the fact that the symmetry factor for the vertex $\phi^k$
has the specific value $1/k!$, as generated by the expansion in
Eq.~(\ref{eq-Sint}), was crucial for the success of these
cancellations.

\begin{figure}
\begin{center}
\centerline{\epsfig{file=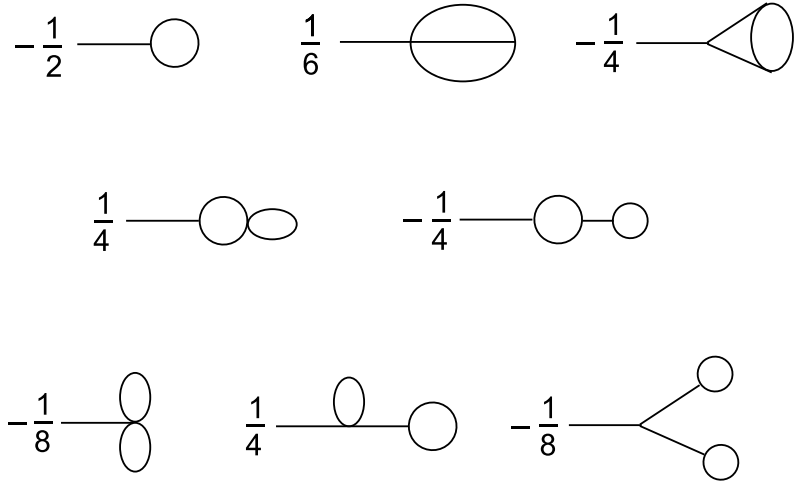,height=6.4cm}}
 \caption{All the diagrams contributing to the mean field
 $\lan\phi(\bm{x})\ran$ up to $\mcal{O}(\sigma^3)$. The first
 line illustrates the three diagrams which survive in the net
 result, cf. Eq.~(\ref{eq-phi11}). The diagrams in the second line
 cancel with each other, and similarly for those in the
 third line.} \label{P1P}
\end{center}
\end{figure}

It is furthermore interesting to notice that the two--loop
contribution to the average field $\lan\phi(\bm{x})\ran$, which is
of $\mcal{O}(\sigma^3)$, is UV--finite as well --- all the
contributing tadpoles mutually cancel, as shown in Fig. \ref{P1P}.
We display here the net result to this order, since this will be
useful later on (see Fig. \ref{P1P} for the respective diagrams) :
 \beq\label{eq-phi11}
 \lan \phi(\bm{x})\ran&\,=\,&-
 \frac{\sigma}{2}\,D(0)+
 \frac{\sigma^3Q_0^4}{6}
 \int_{\bm{z}_{1},\bm{z}_2}\,D(\bm{x}-\bm{z}_1)
 D^3(\bm{z}_1-\bm{z}_2)\nn&{}&
 - \frac{\sigma^3Q_0^6}{4}
 \int_{\bm{z}_{i}}\,D(\bm{x}-\bm{z}_1)
 D(\bm{z}_1-\bm{z}_2) D(\bm{z}_1-\bm{z}_3)
 D^2(\bm{z}_2-\bm{z}_3)\,+\,
 \mcal{O}(\sigma^5)\,.
 \eeq
The apparent $\bm{x}$--dependence of the
$\mcal{O}(\sigma^3)$--terms is only illusory: by homogeneity, the
results of the integrations over $\bm{z}_2$ in the first such a
term, respectively over $\bm{z}_2$ and $\bm{z}_3$ in the second
one, are independent of $\bm{z}_1$; hence, the integral over
$\bm{z}_1$ can be explicitly done with the help of
Eq.~(\ref{eq-intD0}), and then the $\bm{x}$--dependence disappears
indeed.

It becomes more and more tedious to extend such explicit
calculations to higher orders in $\sigma$, and we shall not
attempt to do so. However, we are confident that the pattern of
tadpole cancellations demonstrated by the explicit examples above,
as well by those those to follow in the next subsection, is
non--trivial enough not to be accidental, but rather it is
representative for similar cancellations taking place to all
orders. Based on that, we conjecture that all the $n$--point
functions of $\phi$ with $n\ge 2$ come out finite to all orders in
perturbation theory in this effective theory. Moreover, it seems
that even for the 1--point function $\lan\phi\ran$, the
perturbative corrections are finite beyond one--loop order: the
only UV divergence in this theory seems to be the lowest order
(one--loop) contribution to the mean field, cf.
Eq.~(\ref{eq-phi1}). From Sect. \ref{SECT_Local}, we recall that
$\lan\phi\ran$ enters the definition of the `average saturation
momentum' in the sense of Eq.~(\ref{eq-QSdef1}): $\bar Q_s^2\equiv
Q_0^2\rme^{\sigma\lan\phi\ran}$. Hence, by choosing a physical
value for the latter, one could in principle fix the value of the
ultraviolet cutoff in the effective theory. However, this appears
to be superfluous within the present context, where $\bar Q_s^2$
plays no special role and thus needs not be introduced. Rather, it
is natural to define the saturation momentum in terms of the
vertex operator, whose correlation functions will be discussed in
the next subsection.

\subsection{The $n$--point functions of the vertex operator}
\label{SECT_CorrelV}

Consider now the $n$--point functions of the vertex operator
$V(\bm{x})\equiv {\rme}^{\sigma\phi(\bm{x})}$ --- hence, of the
saturation momentum $Q_s^2(\bm{x})\equiv Q_0^2 V(\bm{x})$ ---,
which are defined as
    \beq\label{eq-Vn}
 \lan V(\bm{x}_1)V(\bm{x}_2)\cdots
 V(\bm{x}_n)\ran\equiv \,\frac{1}{Z}
 \int D[\phi]\,\,V(\bm{x}_1)V(\bm{x}_2)\cdots
 V(\bm{x}_n)\,\,
 \rme^{-S[\phi]}\,.
 \eeq
It is straightforward to construct Dyson equations relating these
quantities to the correlations functions $\lan
\phi(\bm{x}_1)\phi(\bm{x}_2)\cdots \phi(\bm{x}_n)\ran$ of the
Liouville field, and thus deduce the ultraviolet finiteness of the
former from the corresponding property of the latter, as discussed
in the previous subsection. We have already seen this on the
example of the 1--point function for which Eq.~(\ref{eq-Dyson})
implies $\lan V(\bm{x})\ran=1$. For the 2--point function, one
similarly finds
  \beq\label{eq-Dyson2}
 \grad^2_{\bm{x}}\grad^2_{\bm{y}}\!\lan\phi(\bm{x})
 \phi(\bm{y})\ran +
  \frac{Q_0^4}{\sigma^2}\,\lan V({\bm{x}})V({\bm{y}})
  -1\ran=\,({Q_0^2}+\grad^2_{\bm{x}})\delta^{(2)}({\bm{x}}-{\bm{y}})
  \,.
  \eeq
But since Dyson equations are often formal (precisely because of
ultraviolet divergences), it is still instructive to verify the
ultraviolet--finiteness via some explicit calculations in
perturbation theory. Below, we shall do that up to two--loop order
for the 1--point and the 2--point functions of the vertex
operator. As we shall see, the pattern of tadpole cancellations is
now even richer, because it extends to the additional tadpoles
generated by the expansion of the vertex operators.

\vspace*{0.2cm} {\bf (a) One--loop order}

\begin{figure}
\begin{center}
\centerline{\epsfig{file=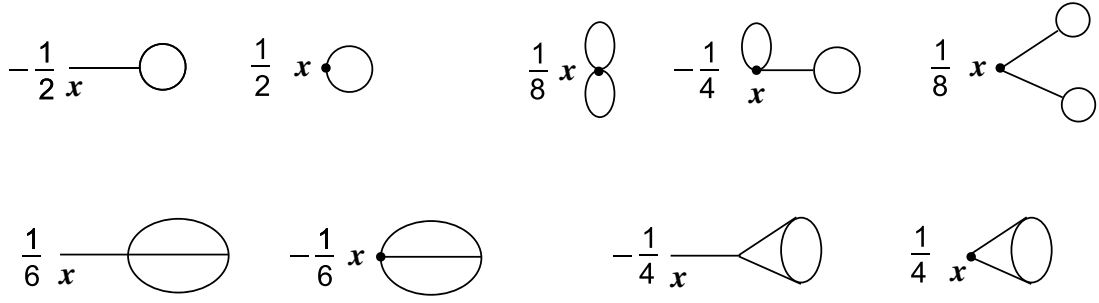,height=4.cm}}
 \caption{All the diagrams contributing to the 1--point function
 $\langle V_{\bm{x}}\rangle$ up to two--loop order. Diagrams within
 a same group cancel with each other.}
\label{V1P2L}
\end{center}
\end{figure}

For the 1--point function $\lan V(\bm{x})\ran$, the corresponding
calculation is straightforward (as before, an upper script on an
expectation value indicates the order in $\sigma$ to which that
expectation value must be evaluated) :
 \beq\label{eq-V11}
 \lan V(\bm{x})\ran&\,=\,&1+\sigma\lan \phi(\bm{x})\ran^{(1)}
 +\frac{\sigma^2}{2}\lan \phi^2(\bm{x})\ran^{(0)}+\,\dots\nn
 &\,=\,&1 - \frac{\sigma^2}{2}\,D(0)+\frac{\sigma^2}{2}\,D(0)
 +\,\dots \,=\,1+\mcal{O}(\sigma^4)\,,\eeq
where we have also used Eq.~(\ref{eq-phi1}). But although very
simple, the above calculation illustrates an important point which
is generic: when computing the $n$--point functions of $V$,
tadpoles generated by the interaction vertices from the action
--- in Eq.~(\ref{eq-V11}), the trilinear vertex which is implicit
inside $\lan \phi\ran^{(1)}$, cf. Fig. \ref{P1P} --- cancel
against other tadpoles generated by the vertices produced when
expanding $V$ --- in Eq.~(\ref{eq-V11}), the quadratic vertex
explicit in the second term there (see also Fig. \ref{V1P2L}).

\begin{figure}
\begin{center}
\centerline{\epsfig{file=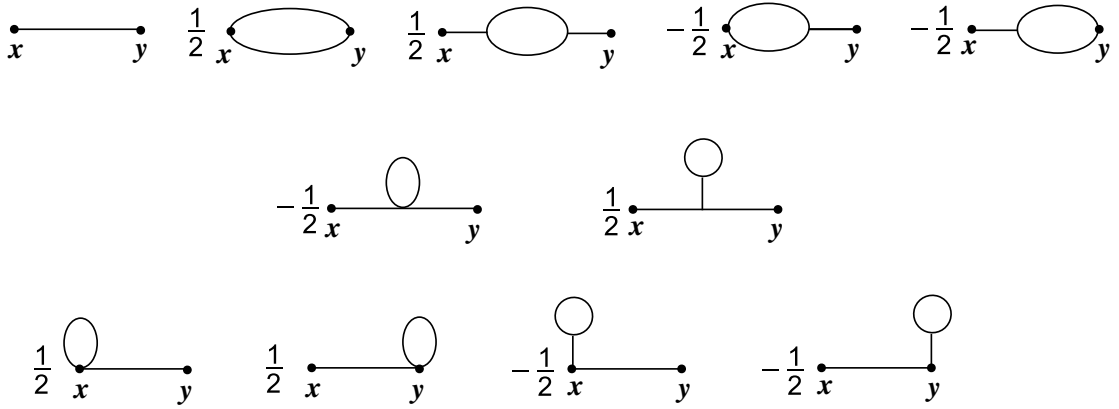,height=5.5cm}}
 \caption{All the diagrams contributing to the 2--point function
 $\langle V_{\bm{x}}V_{\bm{y}}\rangle$ up to $\mcal{O}(\sigma^4)$.
 A vertex with a blob represents a factor of $\sigma$ coming
 from the expansion of the exponential operators.
 The diagrams in the first line are finite and
 contribute to the net result, Eq.~(\ref{eq-V22}).
 The other diagrams are divergent but cancel with each other.
 } \label{V2P1L}
\end{center}
\end{figure}

For the 2--point function, one needs to work a little harder. Note
first that, to $\mcal{O}(\sigma^2)$, $\langle
V_{\bm{x}}V_{\bm{y}}\rangle$ is automatically finite in the
present theory --- at variance to what happens in the free theory,
cf. Eq.~(\ref{eq-QSweak0}) ---, because the only divergences that
could appear to that order are those associated with the
renormalization of the individual vertex operators, but these
cancel out according to Eq.~(\ref{eq-V11}). This is a general
property: since we expect $\lan V\ran=1$ to all orders, in the
calculation of $\langle V_{\bm{x}}V_{\bm{y}}\rangle$ we can
discard all the diagrams contributing to the disconnected piece
$\langle V_{\bm{x}}\rangle \langle V_{\bm{y}}\rangle$,
anticipating that they sum up to one. Then, to
$\mcal{O}(\sigma^4)$ --- corresponding to one--loop order for the
connected piece --- one finds (with $V_{\bm{x}}\equiv V(\bm{x})$,
etc)
 \beq\label{eq-V20}
  \langle V_{\bm{x}}V_{\bm{y}}\rangle
  &\,=\,&1+\sigma^2\left(D_{\bm{x}\bm{y}}+
  \lan \phi_{\bm{x}}\phi_{\bm{y}}\ran^{(2)}\right)
  +\frac{\sigma^3}{2}\lan \phi^2_{\bm{x}}\phi_{\bm{y}}+
  \phi_{\bm{x}}\phi^2_{\bm{y}}\ran^{(1)}\nn&{}&
  +\frac{\sigma^4}{3!}\lan \phi^3_{\bm{x}}\phi_{\bm{y}}+
  \phi_{\bm{x}}\phi^3_{\bm{y}}\ran^{(0)}
   +\frac{\sigma^4}{4}\lan \phi^2_{\bm{x}}\phi^2_{\bm{y}}
   \ran^{(0)}
  +\,\dots
  \eeq
where one should keep only the `connected' pieces of the various
correlation functions appearing in the r.h.s., that is, the
contributions in which the two external points $\bm{x}$ and
$\bm{y}$ are connected with each other. E.g., in
 \beq\label{eq-phi221}
 \lan \phi^2_{\bm{x}}\phi^2_{\bm{y}}
   \ran_0\,=\,2D^2(\bm{x}-\bm{y})+D^2(0),\eeq
one must discard the second, divergent, piece, $D^2(0)\equiv
\langle \phi^2_{\bm{x}}\rangle_0\langle \phi^2_{\bm{y}}\rangle_0$,
since this a part of $\langle V_{\bm{x}}\rangle \langle
V_{\bm{y}}\rangle$. With this rule, the other correlations
appearing in Eq.~(\ref{eq-V20}) are evaluated as
 \beq\label{eq-phi31}
 \lan
 \phi^3_{\bm{x}}\phi_{\bm{y}}\ran_0\,=\,3D(0)D_{\bm{x}\bm{y}},
 \eeq
and, respectively,
 \beq\label{eq-phi21}
 \lan \phi^2_{\bm{x}}\phi_{\bm{y}}\ran^{(1)}   &\,=\,&-
  \frac{\sigma Q_0^2}{3!}
 \int_{\bm{z}}\,\lan\phi^2_{\bm{x}}\phi_{\bm{y}}
 \phi^3_{\bm{z}}\ran_0\,=\,-\sigma D(0)D_{\bm{x}\bm{y}}
 -\sigma Q_0^2
 \int_{\bm{z}}\,D^2_{\bm{x}\bm{z}}D_{\bm{z}\bm{y}},
 \eeq
where we have also used Eq.~(\ref{eq-intD0}). The quantities in
Eqs.~(\ref{eq-phi31}) and (\ref{eq-phi21}) represent one--loop
vertex corrections at $\bm{x}$ (there are, of course, similar
corrections at $\bm{y}$; see Fig. \ref{V2P1L}) and involve
divergent tadpoles. However, when inserted into
Eq.~(\ref{eq-V20}), the symmetry factors are such that these
tadpoles cancel with each other.  Once again, one of these
tadpoles has been generated via contractions inside the vertex
operator $V_{\bm{x}}$ (the one in Eq.~(\ref{eq-phi31})) and the
other one, via contractions inside a vertex from the interaction
piece of the action (that in Eq.~(\ref{eq-phi21})).

Finally, Eq.~(\ref{eq-V20}) involves the connected part of the
2--point function $\lan \phi_{\bm{x}}\phi_{\bm{y}}\ran^{(2)}$,
which has been previously shown to be UV finite (this is given by
the third term in the r.h.s. of Eq.~(\ref{eq-phi23})). All the
diagrams contributing to the r.h.s. of Eq.~(\ref{eq-V20}) ---
including the divergent ones which mutually cancel --- are
displayed in Fig. \ref{V2P1L}. By adding the previous results, one
finds the following, finite, result for $\langle
V_{\bm{x}}V_{\bm{y}}\rangle$ to $\mcal{O}(\sigma^4)$ :
 \beq\label{eq-V22}
  \langle V_{\bm{x}}V_{\bm{y}}\rangle
  &\,=\,&1+\sigma^2 D_{\bm{x}\bm{y}}
 +\frac{1}{2}\big[\sigma^2D_{\bm{x}\bm{y}}\big]^2\nn
  &{}&
  +\frac{\sigma^4Q_0^4}{2}
 \int_{\bm{z}_{1},\bm{z}_2}\,D(\bm{x}-\bm{z}_1)
 D^2(\bm{z}_1-\bm{z}_2)
 D(\bm{z}_1-\bm{y})\nn &{}&
 -   \frac{\sigma^4Q_0^2}{2}
 \int_{\bm{z}}\,\big[D^2_{\bm{x}\bm{z}}D_{\bm{z}\bm{y}}+
 D_{\bm{x}\bm{z}}D^2_{\bm{z}\bm{y}}\big]
  +\,\mcal{O}(\sigma^6).
  \eeq
All the terms in the r.h.s. of Eq.~(\ref{eq-V22}) except for the
last one can be recognized as the expansion of $\exp\{\sigma^2
G_{\bm{x}\bm{y}}\}$, with $G_{\bm{x}\bm{y}}\equiv \lan
\phi(\bm{x})\phi(\bm{y})\ran$ the {\em full} propagator, to the
order of interest. However, the presence of the last term, which
represents vertex corrections, shows that such a simple
exponentiation does not hold in the present theory, in contrast to
what happens in the free theory (recall Eq.~(\ref{eq-QSweakR})).

\comment{ To summarize, the perturbative contributions to
Eq.~(\ref{eq-V20}) of the self--energy type are by themselves
finite, as already demonstrated in the previous subsection,
whereas those of the vertex type add together to a finite result:
tadpoles generated by the vertices implicit in $V$ cancel against
those generated by the vertices from $S_{\rm int}$. This pattern
will be now shown to persist to two--loop order.}

\vspace*{0.2cm} {\bf (b) Two--loop order}

Although considerably more involved (especially for the 2--point
function), the two--loop calculations are also more interesting,
in that they show a richer pattern of tadpole cancellations.

We start with the 1--point function, for which `two loops' means
$\mcal{O}(\sigma^4)$. One has
  \beq\label{eq-V12}
 \lan V(\bm{x})\ran^{(4)}&\,=\,&\sigma\lan \phi(\bm{x})\ran^{(3)}
 +\frac{\sigma^2}{2}\lan \phi^2(\bm{x})\ran^{(2)}+
 \frac{\sigma^3}{3!}\lan \phi^3(\bm{x})\ran^{(1)}+
 \frac{\sigma^4}{4!}\lan \phi^4(\bm{x})\ran^{(0)}
 \,,\eeq
where the first two terms are already known, cf.
Eq.~(\ref{eq-phi11}) and, respectively, Eq.~(\ref{eq-phi23}) ---
within the latter, one has to take $\bm{x}=\bm{y}$, thus yielding
 \beq\label{eq-phi24}
 \frac{\sigma^2}{2} \lan \phi^2(\bm{x})\ran^{(2)}&\,=\,&
 \frac{\sigma^4Q_0^4}{4}
 \int_{\bm{z}_{1},\bm{z}_2}\,D(\bm{x}-\bm{z}_1)
 D^2(\bm{z}_1-\bm{z}_2)
 D(\bm{z}_2-\bm{x})
 \,+\,\frac{\sigma^4}{8}\,D^2(0)
 \,,
 \eeq
---, while the remaining two are easily computed as
  \beq\label{eq-phi3}
 \frac{\sigma^3}{3!}\lan \phi^3(\bm{x})\ran^{(1)}
 \,=\,-
 \frac{\sigma^4Q_0^2}{6}
 \int_{\bm{z}}\,D^3(\bm{x}-\bm{z})
 \,-\,\frac{\sigma^4}{4}\,D^2(0)
 \,,
 \eeq
and, respectively,
 \beq\label{eq-phi4}
 \frac{\sigma^4}{4!}\lan \phi^4(\bm{x})\ran^{(0)}
 \,=\,\frac{\sigma^4}{8}\,D^2(0)
 \,.\eeq
The Feynman graphs associated with these various contributions are
illustrated in Fig. \ref{V1P2L}, where the respective
cancellations are also indicated.  By adding all the
contributions, it is clear that
 \beq\label{eq-V13}
 \lan V(\bm{x})\ran&\,=\,&1+\mcal{O}(\sigma^6)\,.\eeq

\begin{figure}
\begin{center}
\centerline{\epsfig{file=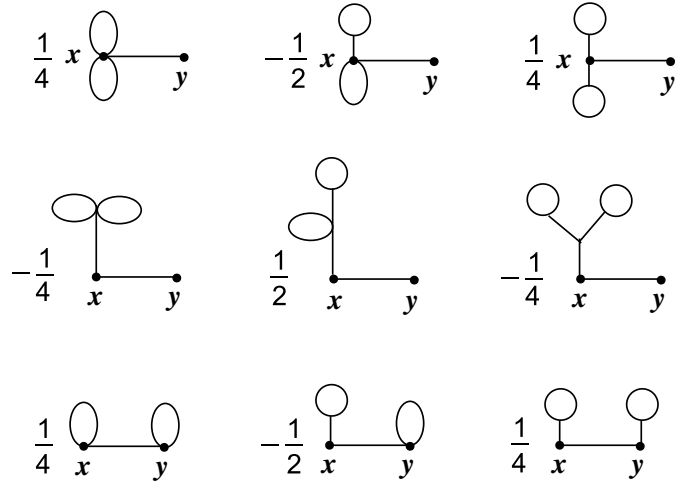,height=6.5cm}}
 \caption{Divergent two--loop diagrams contributing to
 $\langle V_{\bm{x}}V_{\bm{y}}\rangle$ which involve
 double tadpoles. The net result is zero on each of the
 three lines.}
\label{V2P2Tad}
\end{center}
\end{figure}

Consider now the two--loop contributions to $\langle
V_{\bm{x}}V_{\bm{y}}\rangle$, which count to $\mcal{O}(\sigma^6)$.
One of these contributions is $\sigma^2\lan
\phi(\bm{x})\phi(\bm{y})\ran^{(4)}$, which has been already argued
to be UV--finite in Sect. \ref{SECT_CorrelPhi}. The other ones
involve various vertex corrections, which separately develop a
large number of (single or double) tadpoles, but which add
together to a finite result. The respective cancellations show a
pattern which is similar to, but richer than, the one already
observed for the 2--loop self--energy in Sect.
\ref{SECT_CorrelPhi}. In Fig. \ref{V2P2Tad}, we display the
diagrams involving double tadpoles (together with the
corresponding symmetry factors), grouped in such a way to
illustrate the various cancellations. Note that there are three
types of such diagrams, each of them involving three graphs whose
symmetry factors are such that they mutually cancel. We shall not
display the diagrams involving a single tadpole which cancel among
each other, nor the finite ones which yield the net result at this
order, since these diagrams are quite numerous and the result is
not particularly illuminating. Suffices to say that the respective
UV--finite diagrams exhibit the same two--loop topologies as the
self--energy diagrams shown in Fig. \ref{P2P2L}, but which now
appear also in the form of vertex corrections.

\section{Conclusions}
\setcounter{equation}{0}\label{SECT_Conclusion}

In this paper, we have proposed an effective scalar field theory
for the distribution of the saturation momentum in the
two--dimensional impact--parameter space, in QCD at high energy
and large $N_c$. This theory has been constructed as the minimal
field--theoretical generalization of the $\bm{b}$--independent
distribution in Refs. \cite{IMM04,IT04} which is local (in both
$\bm{b}$ and the rapidity $Y$) and which is consistent with the
uncertainty principle and the conformal symmetry of the
high--energy evolution in perturbative QCD to leading order.

The effective action consists in two pieces: one which is
universal (in the sense that it is uniquely fixed by our general
assumptions) and is recognized as the conformally--invariant
Liouville action, and the other which explicitly breaks down
conformal symmetry, thus introducing a physical scale for the
saturation momentum, and which is less universal --- in the sense
that, first, there are many technical options for breaking down
conformal symmetry (here, we have chosen a particularly simple
one: a source term linear in $\phi$) and, second, the structure of
this piece also depends upon the details of the average gluon
distribution that one needs to reproduce. In its minimal version,
the theory is characterized by only two parameters, which are
energy--dependent and homogeneous: the expectation value
$Q_0^2(Y)\equiv \langle Q_s^2\rangle$ of the saturation momentum
and the coupling constant $\sigma(Y)$ which characterizes the
disorder introduced by fluctuations, and corresponds to the `front
dispersion' in the previous literature \cite{IMM04,IT04}. But
additional parameters can be added, if needed (e.g., in order to
describe a inhomogeneous distribution), via the symmetry--breaking
term.

The description offered by this effective theory is certainly
crude and oversimplified as compared to the original QCD problem.
It has some obvious shortcomings --- e.g., it cannot accommodate
the long--range correlations in impact--parameter space, which are
non--local in rapidity --- and there might be some other, less
obvious, ones, which are however difficult to recognize in the
absence of explicit QCD calculations (like solutions to the
Pomeron loop equations). On the other hand, we expect this theory
to be correct at least at short distances, of the order of the
average saturation length $1/Q_0(Y)$ or smaller, and this is
important, since this is precisely the range where the predictions
of the theory are more interesting, and also universal, since
determined by the Liouville piece of the action alone.

Specifically, there are two interesting predictions: \texttt{(i)}
The short--range correlations are power--like, with the powers
determined by the natural dimension of the squared saturation
momentum $Q_s^2$, which is $2$. This result is a direct
consequence of the conformal symmetry, and it is indeed natural to
find manifestations of this symmetry on distance scales which are
shorter than any symmetry--breaking length scale in the problem
(here, $1/Q_0$). \texttt{(ii)} The fluctuations in the saturation
momentum are pushed to much harder scales than expected from
previous analyses: the minimum of the potential, corresponding to
the most probable value for $Q_s^2$, occurs for $Q_s^2=\langle
Q_s^2\rangle \equiv Q_0^2 \langle {\rm e}^{\sigma\phi}\rangle$,
rather than for $Q_s^2=\bar Q_s^2\equiv
Q_0^2\,{\rme}^{\sigma\langle\phi\rangle}$, as it was the case for
the coarse--grained distribution in Eq.~(\ref{eq-prob}). This is a
reflection of the uncertainty principle, together with the
multivalence of the saturation momentum: $Q_s^2(\bm{x})$, which is
an operator in the present theory, is also the measure of the size
of the fluctuation at $\bm{x}$, and hence the natural value for
its gradient. Whereas in the context of the coarse--grained
approximation (\ref{eq-prob}), the value $\langle Q_s^2\rangle$
corresponds to a very rare fluctuation, in the tail of the
distribution, this is not so anymore in the context of the
effective field theory, where fluctuations with arbitrarily high
values for $Q_s^2$ are allowed, since they have tiny sizes and
thus give a small contribution to the action.

It would be very interesting to understand the physical
consequences of these new results. To that aim, the formalism
needs to be further developed, to allow for the calculation of
observables. A possible direction in that sense will be sketched
in the Appendix, where we show --- within the simple context of
the McLerran--Venugopalan model --- how to couple the present
field theory describing the statistics of the saturation momentum
to the CGC formalism, where the small--$x$ gluons are represented
by a color charge density in terms of which observables can be
constructed in a standard way \cite{EDICGC}.

For our particular choice for the symmetry--breaking source term
(an operator linear in $\phi$), we found that the theory has a
remarkable property: the correlation functions of interest (in
particular, those of the vertex operator which enters the
definition of the saturation momentum) are ultraviolet finite when
computed in perturbation theory, and thus do not call for
renormalization. Our proof in that sense is not complete (it is
limited to two--loop calculations), so it would be interesting to
extend that to an all--order proof, and also to put this result in
a more general perspective --- e.g., to clarify the general
conditions under which a theory of this kind (a deformation of the
Liouville field theory) has UV--finite correlations.

It should be relatively straightforward to perform lattice
simulations for the effective theory that we propose, and thus
better understand its properties in the strong coupling/high
energy regime. The theory is finite and has no special symmetry,
so there is no need for special care in constructing its lattice
regularization. Also, lattice artifacts should rapidly vanish in
the continuum limit. Moreover, it might be possible to
analytically investigate this theory, via non--perturbative
methods: formally, this theory is just Liouville field theory
perturbed by a relatively simple operator --- a source term linear
in $\phi$.

\section*{Acknowledgments}

This paper had a rather lengthy gestation with several critical
phases during which we have profited from discussions with
numerous colleagues. We acknowledge useful conversations with
Kazunori Itakura, Andrei Parnachev, and Gr\'egory Soyez,
insightful observations from Kenji Fukushima and Yoshitaka Hatta,
and patient explanations and critical remarks from Al Mueller.
Both authors are grateful to the INT, Washington University,
Seattle and to the organizers of the program ``From RHIC to LHC:
Achievements and Opportunities'' for hospitality during one of the
critical phases alluded to above.

This manuscript has been authorized under Contract No. DE-
AC02-98CH0886 with the US Department of Energy.

\appendix
\section{The random McLerran--Venugopalan
model}

Given the effective theory for $\phi$, which describes the
fluctuations in the saturation momentum, one can wonder how to
compute the effects of these fluctuations on physical observables,
so like the scattering amplitude of a projectile dipole. To that
aim, one can resort on the CGC formalism, in which the observables
are expressed in terms of the color charge density which
represents the source of the gluon distribution. Using the
Liouville field $\phi$, it is possible to couple this color charge
density to the fluctuating saturation momentum in a
conformally--invariant way.  As a simple example, let us describe
a `random' extension of the McLerran--Venugopalan model \cite{MV},
which is conformally--invariant. Recall that, in this model, the
color charge density in a large nucleus, $\rho_a(\bm{x})$, is a
Gaussian random field variable with zero expectation value and
local 2--point function proportional to the saturation momentum:
  \beq\label{eq-MV}
 W_{\rm MV}[\rho]= \mcal{N}\exp\left\{-\int \dif^2 \bm{x}\,
\frac{\rho_a(\bm{x})\rho_a(\bm{x})}{2Q_s^2(\bm{x})}\right\}\,.\eeq
($\mcal{N}$ is a normalization constant such that $\int\!
D[\rho]\,W_{\rm MV}[\rho]=1$.) In the original formulation of the
model, $Q_s^2(\bm{x})$ is a given (generally, point--dependent)
quantity, which fixes the strength of the color charge density
squared, and hence of the gluon distribution. In the present
context,  this is interpreted as
$Q_s^2(\bm{x})={Q_0^2}\,{\rme}^{\sigma\phi(\bm{x})}$, with
$\phi(\bm{x})$ a random field distributed according to the
effective theory. This leads to a statistical field theory for the
coupled fields $\phi$ and $\rho_a$, with the following partition
function
 \beq\label{eq-MVR}
 Z=   \int\!D[\phi] \int\! D[\rho_a]\,
\exp\left\{-\int \dif^2 \bm{x}\,\left(
 \frac{1}{2}(\grad^i \phi)^2 +
 \frac{Q_0^2}{\sigma^2}\,{\rme}^{\sigma\phi}\,+\,
\frac{\rho_a\rho_a}{2Q_0^2}\,{\rme}^{-\sigma\phi} \right
 )\right\}\,,\eeq
which is conformal invariant, as anticipated. To see this, note
that under the scale transformation in Eq.~(\ref{eq-lambda}), the
field $\rho_a(\bm{x})$, which has mass dimension two, transforms
as follows:
  \beq\label{eq-rholambda}
 \bm{x} \,\to\,\bm{x}' = \lambda \bm{x}\quad\Longrightarrow\quad
 \rho_a(\bm{x})&\,\to\,&
 \rho_a'(\bm{x}')\,=\, \frac{1}{\lambda^2}\,\rho_a(\bm{x})\,, \eeq
so that the term coupling $\rho$ to $\phi$ in the action in
Eq.~(\ref{eq-MVR}) is indeed invariant, so like the other terms
there. Of course, this symmetry is ultimately broken by the source
terms (for either $\rho$ or $\phi$), which need to be added to the
action in Eq.~(\ref{eq-MVR}) in order to generate a non--trivial
expectation value for the saturation momentum (or for the local
gluon distribution).

%\bibliographystyle{unsrt}
%\bibliography{myrefs}

\end{document}